\documentclass{elsart}

\usepackage{epsfig}

\usepackage{amssymb}
\usepackage{subfigure}

\begin{document}

\begin{frontmatter}

\title{Temperature characterization of scintillation detectors using solid-state photomultipliers for radiation monitoring applications}

\author[sgchiram]{Clarisse Tur, }
\author[sgchiram]{Vladimir Solovyev}
\author[sgcnemours]{, and J\'er\'emy Flamanc}
\corauth[cor1]{Corresponding Author: clarisse.tur@saint-gobain.com,
(440)834-5706, fax (440)834-7683}
\address[sgchiram]{\scriptsize{Saint-Gobain Crystals, 17900 Great Lakes Parkway, Hiram, OH 44234, USA}}
\address[sgcnemours]{\scriptsize{Saint-Gobain Cristaux et D\'etecteurs, B. P. 521, 77794 Nemours Cedex, France}}

\begin{abstract}

We have characterized two state-of-the-art solid-state photomultipliers,
one by SensL, the other by Hamamatsu, coupled to scintillators by
Saint-Gobain Crystals in the  $-25\,^{\circ}\mathrm{C}$ - 
$+50\,^{\circ}\mathrm{C}$ temperature range. At room temperature, the energy
resolution at 661.6 keV measured with both detectors is worse than the 
resolution obtained when the crystals are coupled to a regular photomultiplier
tube. Both the pulse height and pulse height resolution of the 661.6 keV 
gamma rays in the $^{137}$Cs spectrum vary strongly with temperature. The 
noise threshold determined from the $^{22}$Na spectrum increases quadratically 
as the temperature is increased to well above 100 keV at 
$+50\,^{\circ}\mathrm{C}$ for both detectors. 

\end{abstract}
\end{frontmatter}

\section{Introduction}

In recent years, solid-state photomultipliers (SSPMs) have emerged as a 
promising light sensor for compact, low-bias scintillation detectors. 
They have a gain comparable to that of the typical photomultiplier tube (PMT) 
while offering a number of key advantages compared to the latter such as 
compactness, operation at low bias voltage, robustness, insensitivity to 
magnetic fields, and in some cases, fast timing properties. In addition, 
operation under ambient light, while not recommended for optimum noise 
levels, will not cause permanent damage to the SSPM. Sizes are increasing 
as smaller individual sensors are tiled together.

The SSPM is an array of a large number of small pixels (typically a few 
tens of $\mu$m), each consisting of a Geiger-mode avalanche photo-diode 
coupled to a quenching resistor. When used in a scintillation detector, each 
pixel of the SSPM will emit the same saturated signal if hit by an 
optical photon when operated at a reverse bias above the breakdown 
voltage (V$_{br}$) signaling the presence of the photon. The Geiger discharge 
responsible for the signal is stopped when the voltage is brought back to 
below V$_{br}$ by the quenching resistor. Thus each pixel is operated in a 
binary, on/off, mode to indicate the presence or absence of a photon. The 
SSPM is therefore fundamentally a photon counting device where the number 
of ``on'' pixels is proportional to the energy deposited in the scintillator,
so long as the the number of pixels excited is small compared to the total.
In other words, the number of pixels available is the limiting factor for 
the dynamic linear range of the SSPM.

\begin{figure*}
\centering
\centerline{
\mbox{\includegraphics[width=0.497\textwidth]{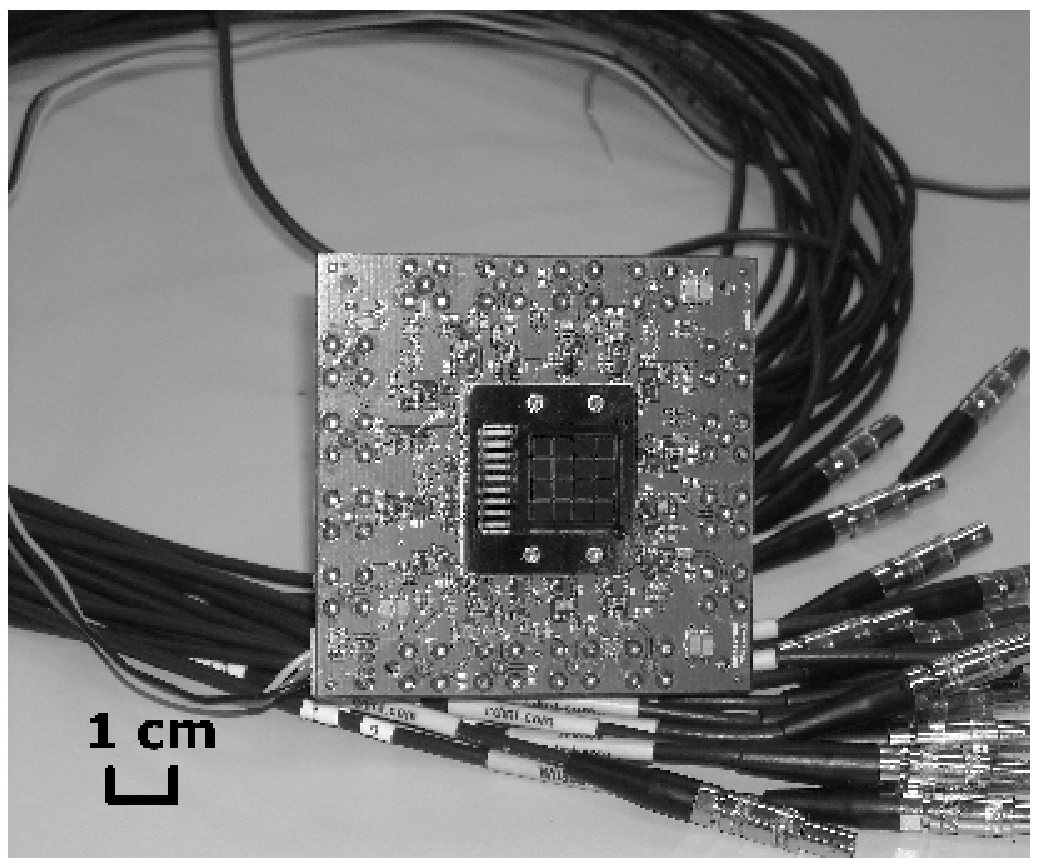}}
\mbox{\includegraphics[width=0.497\textwidth]{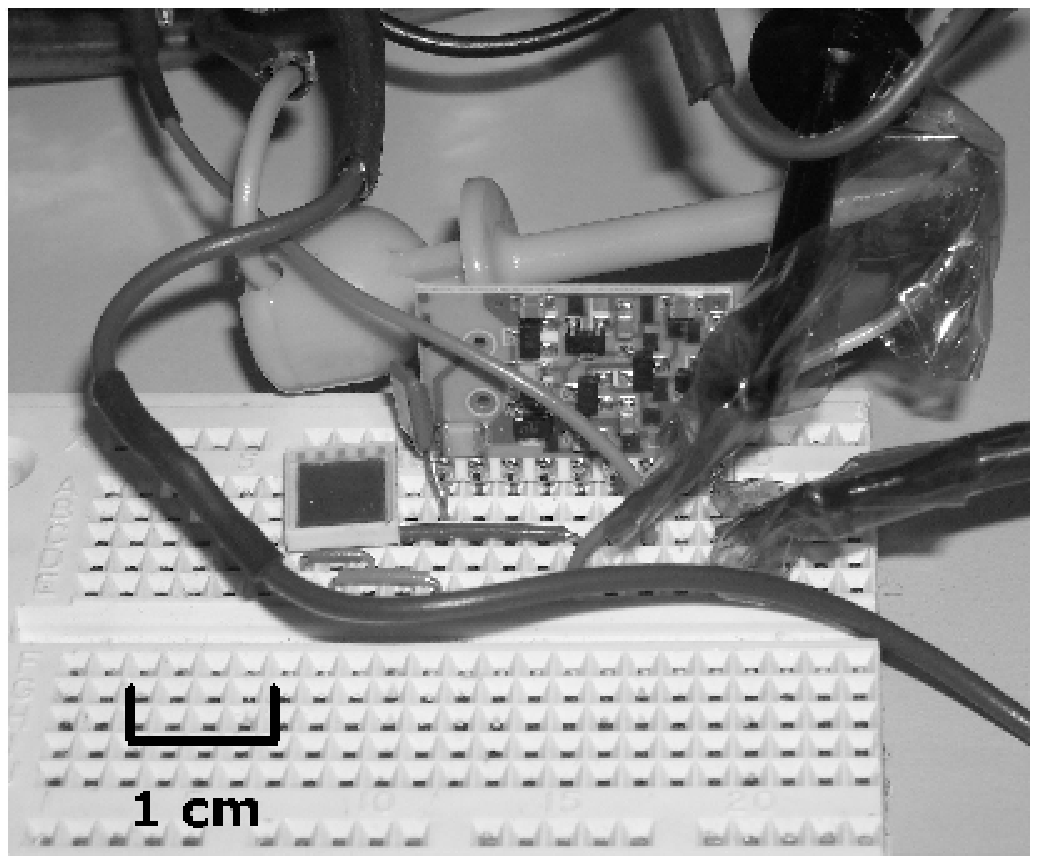}}
}
\centerline{
\mbox{\includegraphics[width=0.497\textwidth]{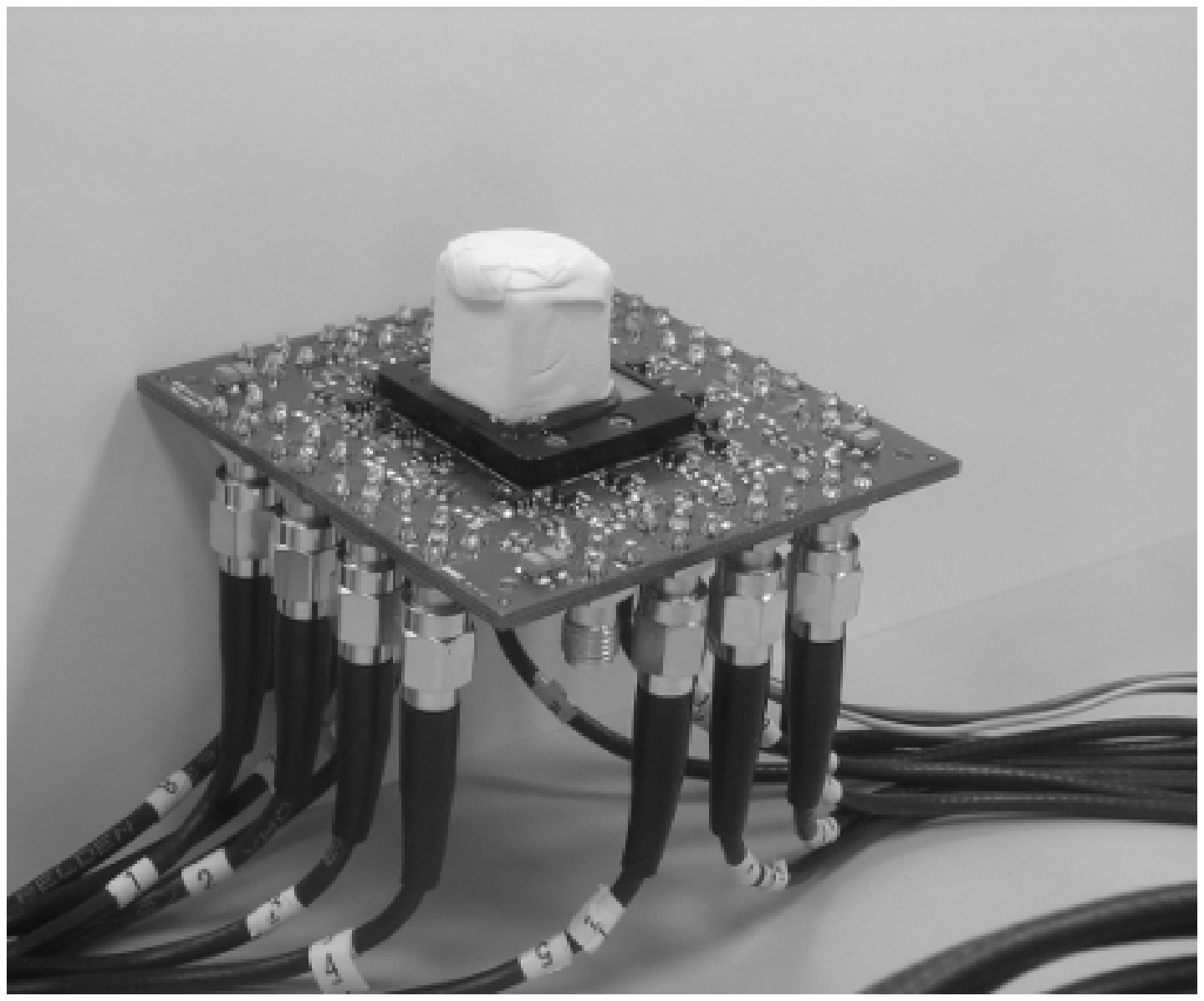}}
\mbox{\includegraphics[width=0.497\textwidth]{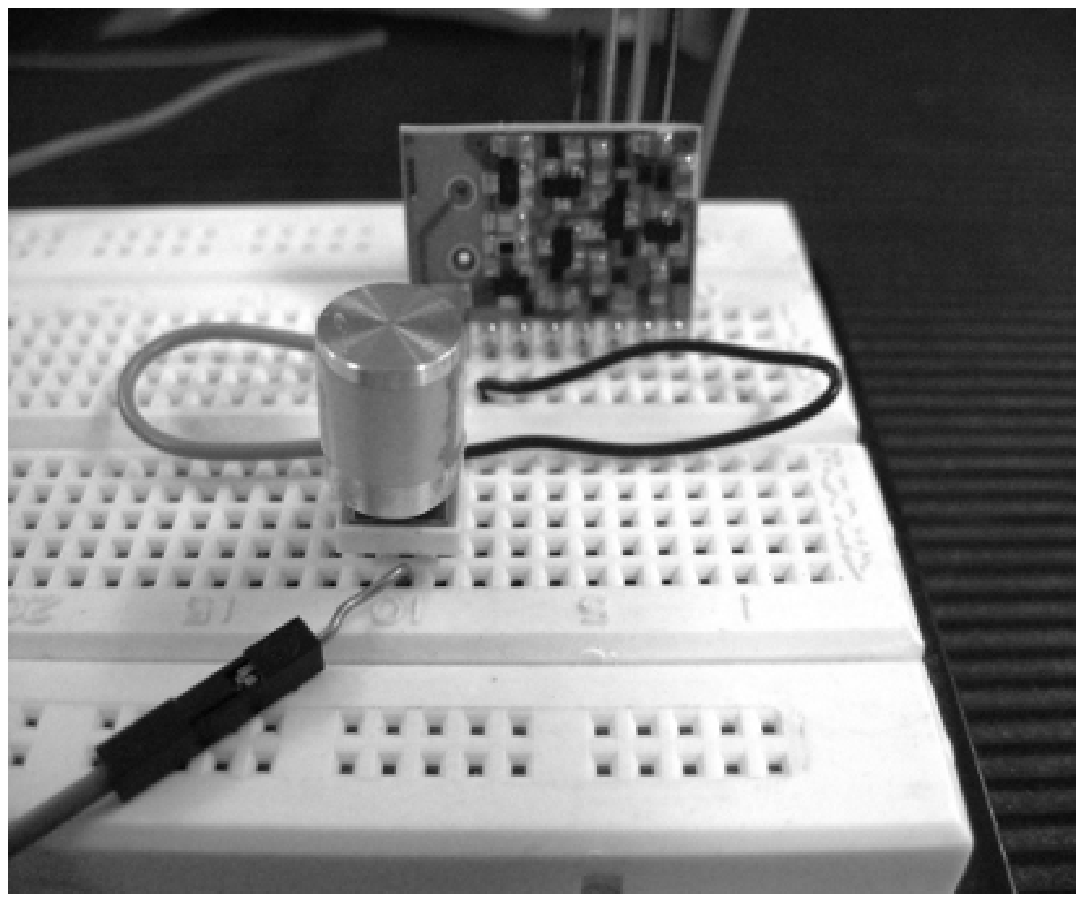}}
}
\caption{\emph{Top-Left:} SensL's 4x4-die SSPM and pre-amplifier/power board. 
16 cables deliver the output signal of each of the 16 SSPM dies. 
\emph{Top-Right:} Hamamatsu's 2x2-die MPPC (Multi-Pixel Photon Counter) 
placed on a breadboard next to the Hamamatsu H4083 photodiode 
pre-amplification board, the 100 M$\Omega$ load-resistor,
and the cables connecting them. The output signal of the 4 MPPC dies are 
summed into 1 signal output. \emph{Bottom-Left:} The SensL SSPM coupled to the 
CsI(Tl) crystal with silicone grease. The crystal is wrapped in white 
reflective Teflon sheet.\emph{Bottom-Right:} The Hamamatsu MPPC coupled to
the BrilLanCe 380 (LaBr$_{3}$(Ce)) crystal with silicone grease. The crystal is 
encapsulated in an aluminum housing with a glass optical window for protection
from humidity.}
\label{fig1}
\end{figure*}

Much progress has been made on the SSPM technology front since the first 
publications on the topic (\cite{aki97}-\cite{bon00}) and a number of papers
have followed since on its various possible applications ranging from the
general subject of radiation detection (\cite{pie06}-\cite{sta07}) to more 
specific applications such as medical imaging (\cite{sta05}-\cite{sch09}),
basic research in physics (\cite{buz03}-\cite{min08}), low-intensity light 
biosensors (\cite{lin05}), and laser radar systems (\cite{mar03}, 
\cite{joh03}).    

In the present paper, we characterize 2 silicon-based SSPMs, one by
SensL (\cite{sen09}), the other by Hamamatsu (\cite{ham09}),  coupled 
to scintillators by Saint-Gobain Crystals (\cite{sgc09}), CsI(Tl) for the
former and LaBr$_{3}$:Ce for the latter. We will use Hamamatsu's trademark 
designation of MPPC (Multi-Pixel Photon Counter) hereafter whenever we 
refer to the Hamamatsu sensor. At the time of 
the tests, the SensL SSPM offered the largest area of a 
single unit detector, while the Hamamatsu MPPC offered the highest density
of Geiger cells among the commercially available sensors of this type.
Beyond the basic performance, our main interest is temperature dependence 
of the two scintillation detectors within the 
$-25\,^{\circ}\mathrm{C}$ - $+50\,^{\circ}\mathrm{C}$ range. To our knowledge, 
such a complete temperature analysis of an SSPM-based
scintillation system has not been performed prior to this study. More 
precisely, we quantified the variations within the afore-mentioned 
temperature range of the noise threshold using a $^{22}$Na source, and of 
the pulse height (PH) and pulse height resolution (PHR) of the 661.6 keV 
peak in $^{137}$Cs.

In Section 2, we describe our scintillation detectors and experimental setups. 
We report the results of our characterization in Section 3, and discuss those
results in Section 4.

\section{Experimental Setup}

\subsection{The SensL SSPM setup}

The SensL SSPM (13 mm x 13 mm) is an array of 16 smaller SSPMs 
(3 mm x 3mm), hereafter referred to as ``dies'', in a 4x4 arrangement. Each 
die is made of 3,640 pixels (35 $\mu$m x 35 $\mu$m), 
resulting in 58,240 pixels for the entire SSPM. The SSPM is pre-installed
on a trans-impedance pre-amplification/power board with 16 output cables 
each delivering the pre-amplified signal of one of the 16 SSPM dies (see 
top-left and bottom-left photos on Figure \ref{setup}). A DC 
power supply is used to provide +5 V, -5 V, and ground connections to the 
pre-amplifier/power board, which itself generates the bias to operate 
the SSPM. This bias can be mechanically adjusted using a screw on the 
afore-mentioned board between the breakdown voltage, V$_{br}$, of 28 V and the
highest recommended voltage of 32 V. We chose to operate the SSPM at the 
manufacturer-recommended bias of 30 V, which is in the middle of the 
recommended range, as it gives the best compromise between photon 
detection efficiency (PDE) and noise. To verify this point, we performed 
a simple measurement where we observed the Full Width at Half Maximum (FWHM) 
of the 661.6 keV peak at room temperature as a function of the bias voltage 
varied between 28 V and 32 V. The value of this FWHM decreases 
(i.e. improves) as the bias is increased from 28 V to 30V, where it 
reaches a minimum, then the FWHM increases (or worsens) as the bias is 
raised further toward 32 V.

The PDE of an SSPM is the probability that a pixel generates a Geiger discharge
upon being hit by a photon and is given by the following equation:
 
\begin{equation}
PDE(\lambda,V)=\eta(\lambda)\times\epsilon(V)\times F
\label{thePDE}
\end{equation}

where $\lambda$ is the wavelength of the incoming photon, V is the bias, 
$\eta$($\lambda$) is the quantum efficiency of the silicon, $\epsilon$(V) 
is avalanche initiation probability, and $F$ is the geometrical efficiency or 
fill factor of the SSPM (\cite{sen09}). Both the PDE and the dark count rate
of the SSPM have been demonstrated (see e.g. \cite{kin98}) to increase as 
the excess bias V$_{ex}$, defined as the difference between the operating 
bias V$_{op}$ and the breakdown voltage V$_{br}$ 
(i.e. V$_{ex}$ = V$_{op}$ - V$_{br}$), is increased. Hence a 
V$_{op}$ has to be selected which offers the best compromise between a
high-enough PDE and a low-enough noise level. 
 
\begin{figure*}
\centering
\centerline{
\mbox{\includegraphics[width=0.497\textwidth]{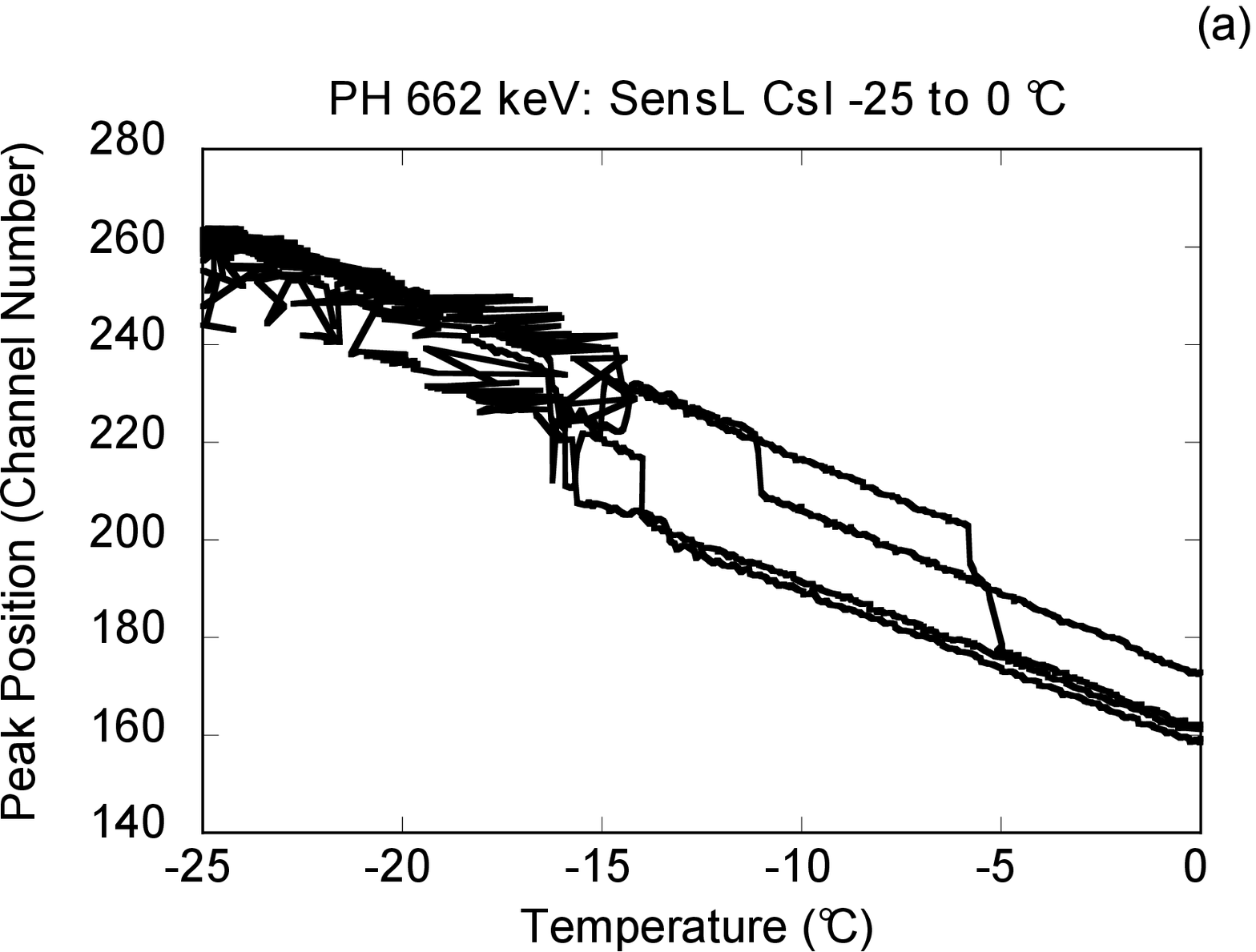}}
\mbox{\includegraphics[width=0.497\textwidth]{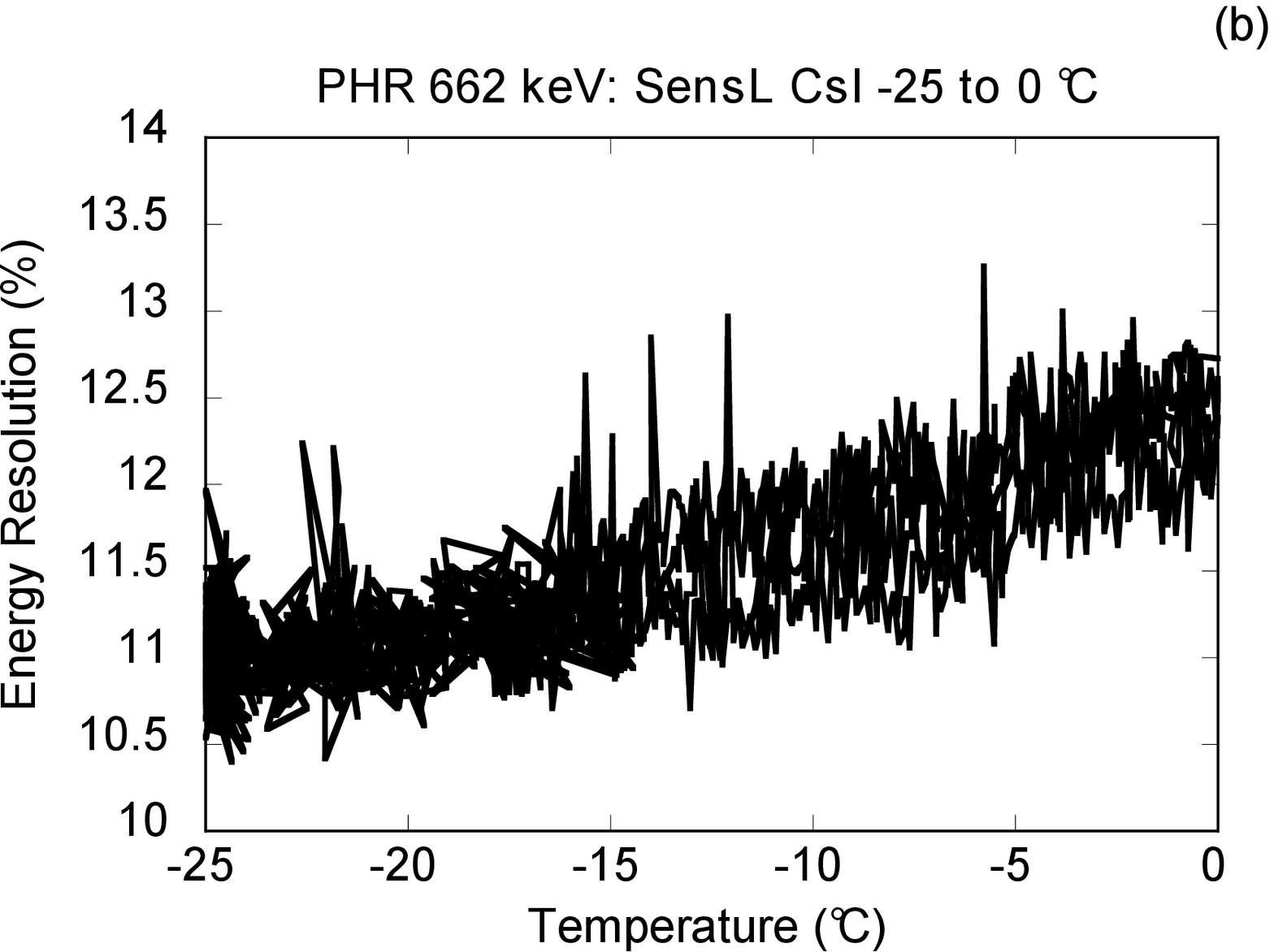}}
}
\centerline{
\mbox{\includegraphics[width=0.497\textwidth]{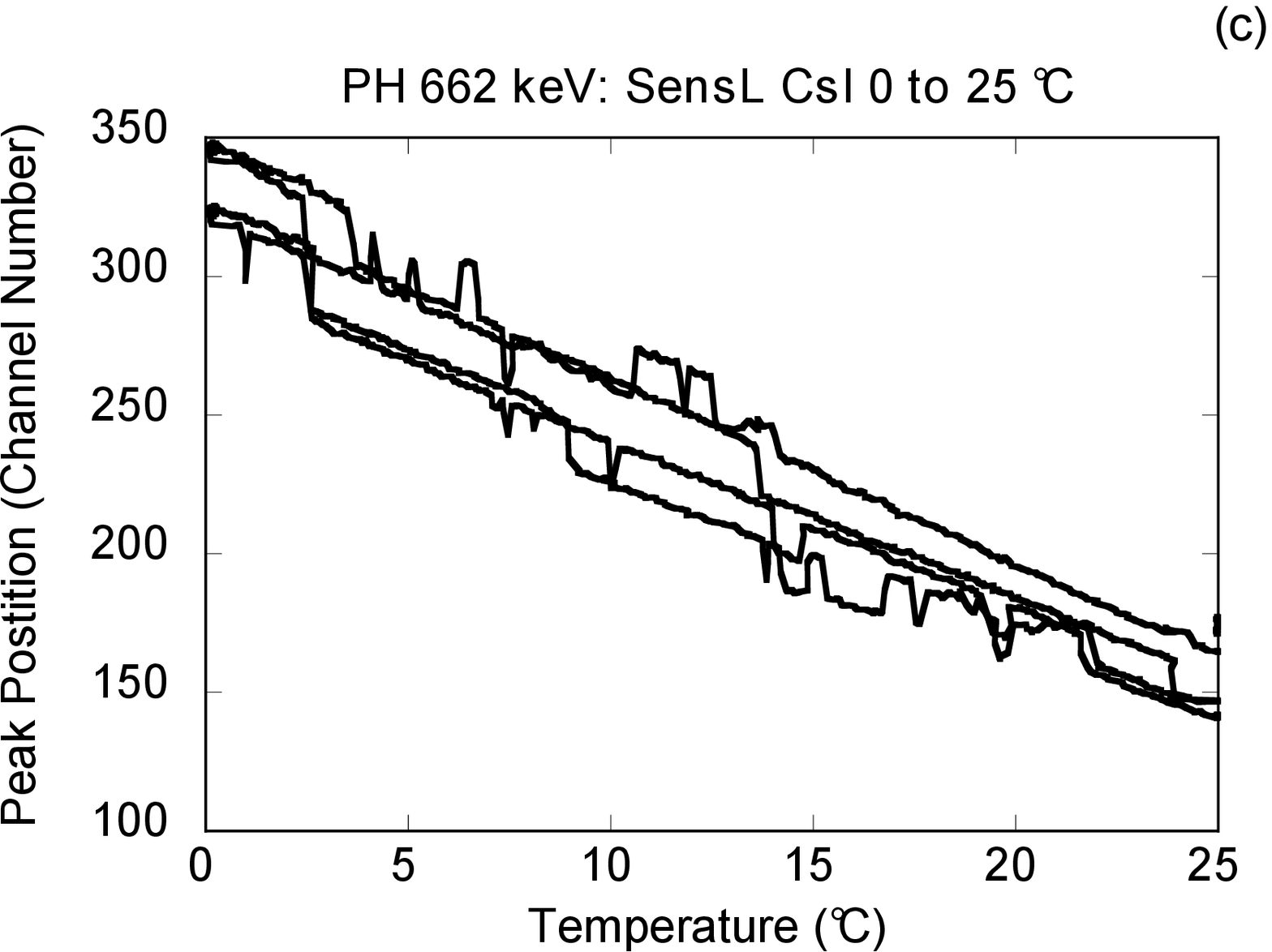}}
\mbox{\includegraphics[width=0.497\textwidth]{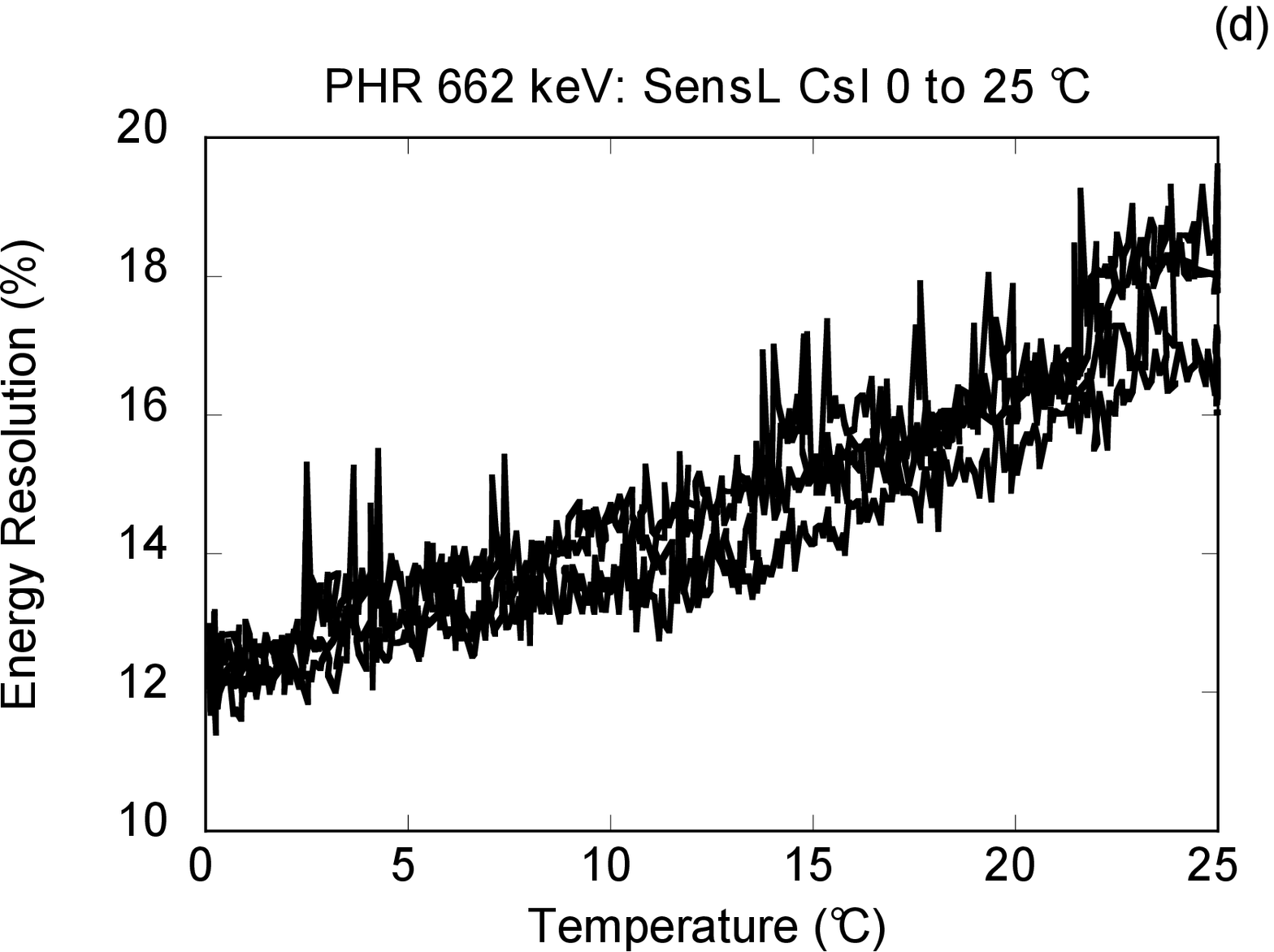}}
}
\centerline{
\mbox{\includegraphics[width=0.497\textwidth]{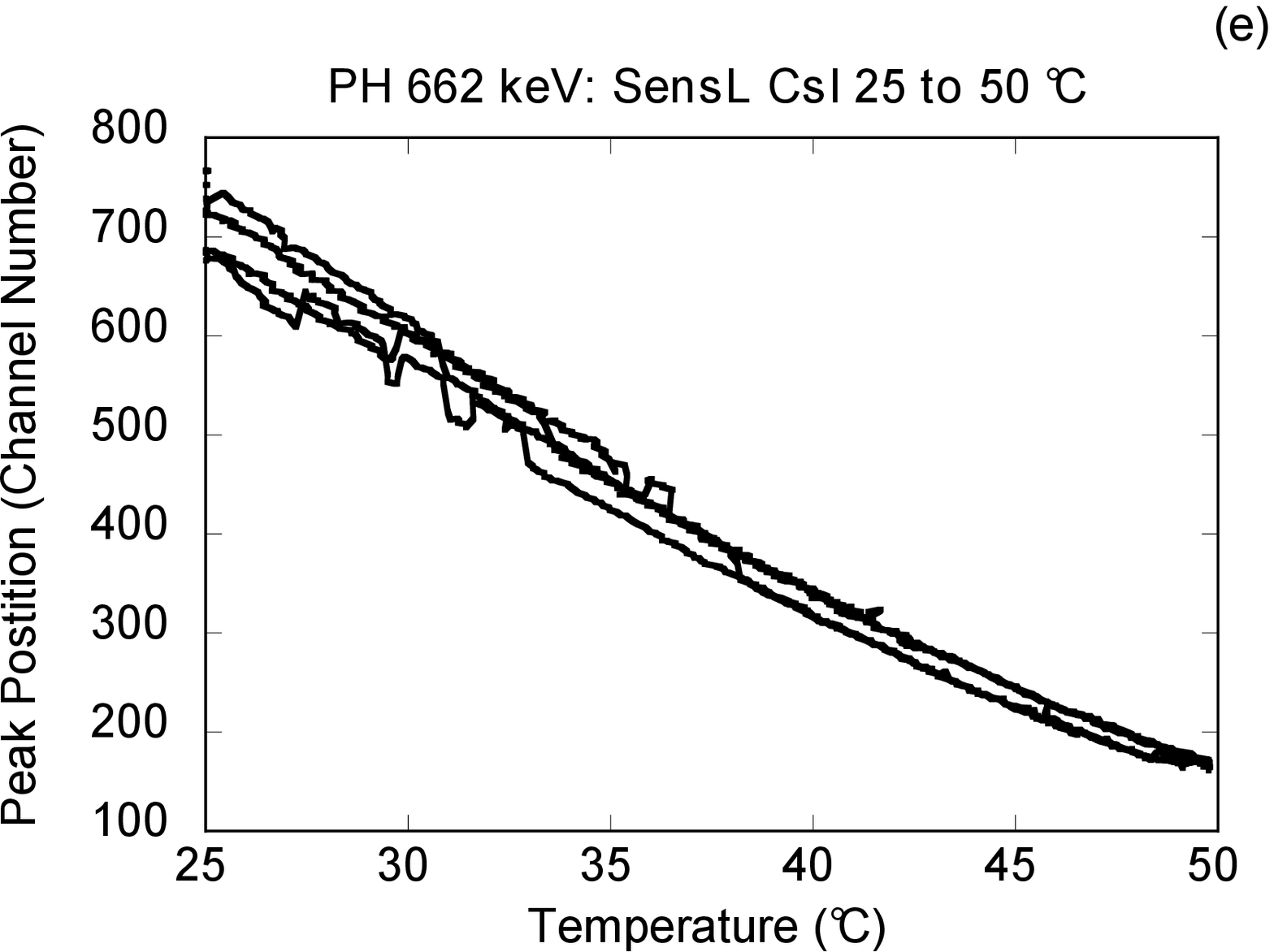}}
\mbox{\includegraphics[width=0.497\textwidth]{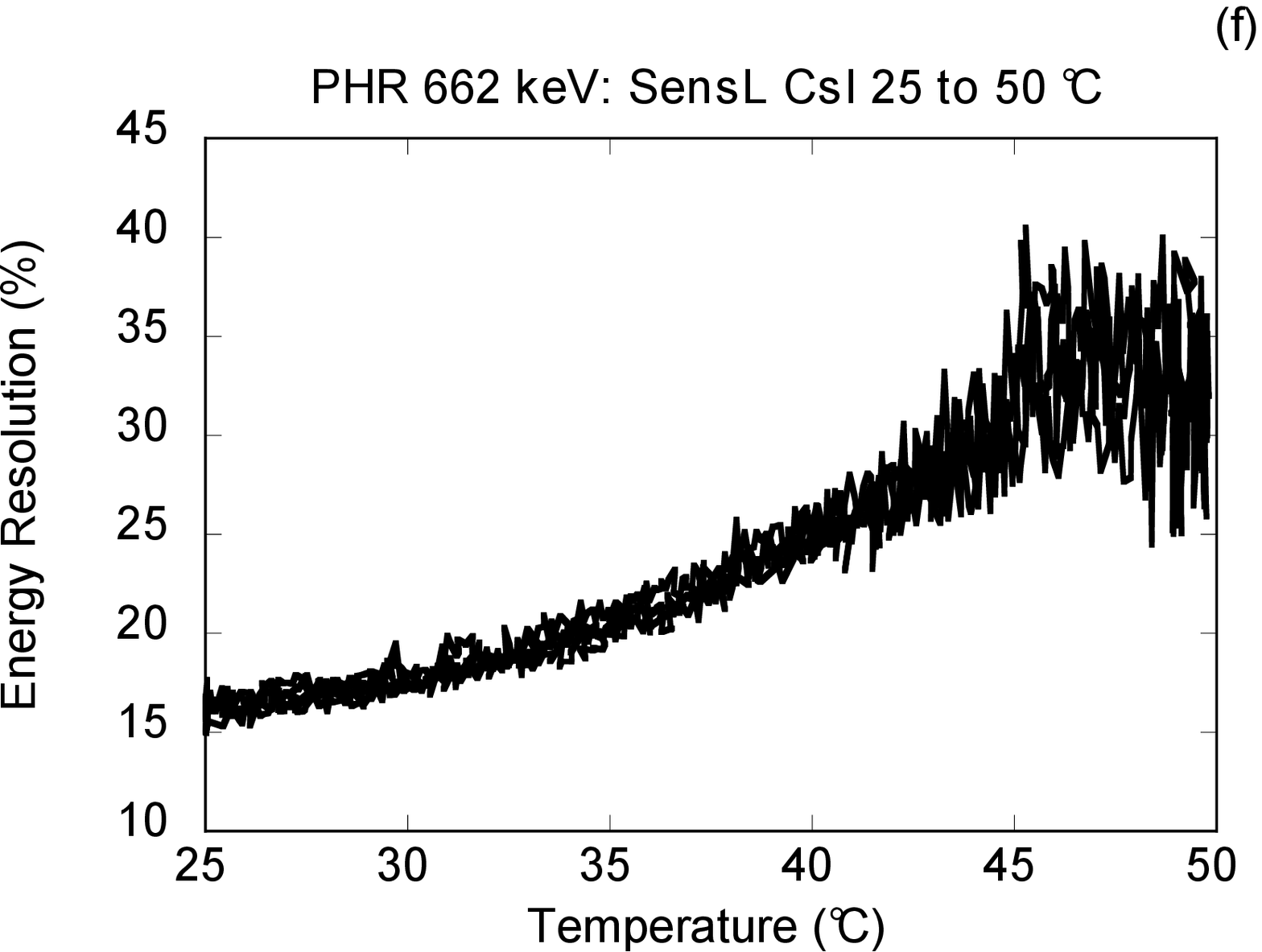}}
}
\caption{Temperature dependence, for the SensL SSPM + CsI(Tl) scintillation 
detector, of (a) the PH from -25 to
$0\,^{\circ}\mathrm{C}$, (b) the PHR from -25 to $0\,^{\circ}\mathrm{C}$,
(c) the PH from 0 to $25\,^{\circ}\mathrm{C}$, (d) the PHR from
0 to $25\,^{\circ}\mathrm{C}$, (e) the PH from 25 to 
$50\,^{\circ}\mathrm{C}$, and (f) the PHR from 25 to 
$50\,^{\circ}\mathrm{C}$ of the 661.6 keV peak in the $^{137}$Cs spectrum.
The amplifier gain settings are different for each of the 3 temperature 
ranges.}
\label{sensLTempDep}
\end{figure*}

The PDE of the SensL SSPM peaks at about 470 nm. We therefore
decided to couple it to a cubic CsI(Tl) scintillator by Saint-Gobain Crystals,
13 mm x 13 mm x 13 mm in size, via silicone grease. The crystal was not 
encapsulated since it is only slightly hygroscopic. The sides not coupled 
to the SSPM were wrapped in white Teflon reflector. The PDE of the SSPM at the 
CsI(Tl) wavelength of maximum emission of 550 nm, when operated at the 
recommended 30 V bias at room temperature, is about 8\% according to SensL. A 
radioactive source was positioned on top of the pre-amplifier/power board, 
which was placed in a black plastic box to ensure light-tightness. The black 
box itself was placed in a temperature test chamber (by Thermotron Industries) 
in which the temperature can be  varied within the $-70\,^{\circ}\mathrm{C}$ - 
$+180\,^{\circ}\mathrm{C}$ range. The moisture level in the temperature test 
chamber was not monitored or controlled. The DC power supplies and the readout 
electronics were located outside the temperature test chamber.

The 16 output signals from the SensL pre-amplifier board were added using 2 
LeCroy 428F linear fan-in-fan-out modules. This summed output was 
amplified by a Canberra 2020 spectroscopy amplifier (with a 3 $\mu$s shaping
time, since this setting gave the best PHR at 661.6 keV at room temperature) 
before being digitized by a Canberra 8075 analog-to-digital converter 
(ADC) module. The resulting output was analyzed and saved by a multi-channel
analyzer (MCA).

\subsection{The Hamamatsu MPPC setup}

\begin{table*}
\centering
\caption{Standard deviation (denoted by $\sigma$) and mean (denoted by $\mu$)
of the measured temperature during soaks 
at constant temperature for 4 temperature settings and the same statistics 
for the 661.6 keV PHR as measured by the SensL SSPM + CsI(Tl) scintillation
detector at those 4 temperatures. The statistics on the temperature are 
in $^{\circ}$C and those on the PHR in \%.}
\label{sensLstats}
\begin{tabular*}{0.95\textwidth}{ccccc}
\hline 
\hline 
Temperature & $\mu$(Temperature) & $\sigma$(Temperature) 
& $\mu$(PHR) & $\sigma$(PHR)  \\
\hline 
-25  &  -24.67  & 0.40  & 11.03  & 0.23 \\
0    &    0.18  & 0.08  & 12.36  & 0.29 \\
25   &   24.99  & 0.09  & 16.84  & 1.19 \\
50   &   49.31  & 0.33  & 32.15  & 2.57 \\
\hline
\end{tabular*}
\end{table*}

The Hamamatsu MPPC (6 mm x 6 mm) is an array of 4 dies (3 mm x 3 mm) in a 
2x2 arrangement. Each die consists of 14,400 pixels (25 $\mu$m x 25 $\mu$m) 
without any dead space between the 4 dies, resulting in 57,600 pixels for the 
entire MPPC. The recommended V$_{op}$ is about 70 V; we chose to bias the
MPPC at 65 V (for reasons which will become clear in the next Section) through
a load resistor of 100 M$\Omega$.

\begin{figure}
\centering
\includegraphics[width=0.8\textwidth]{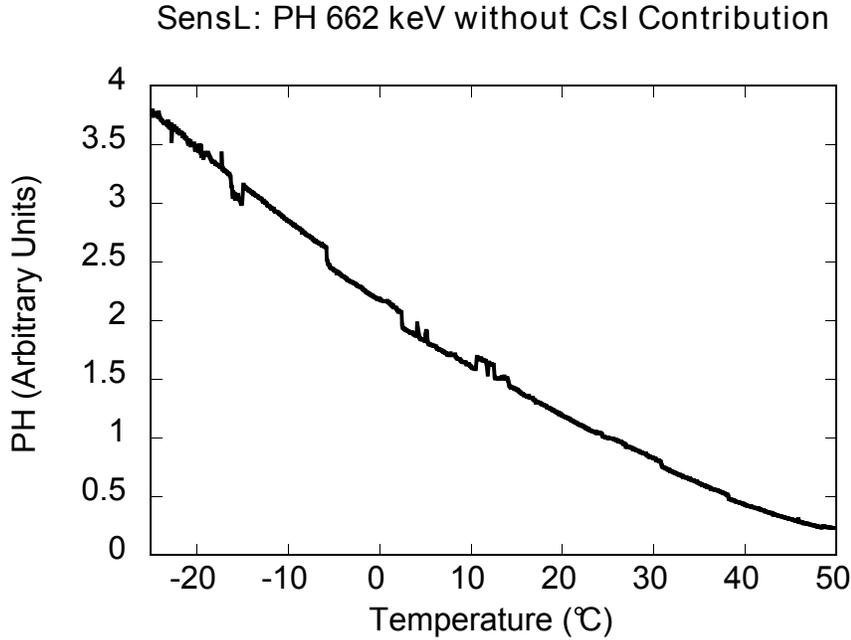}
\caption{PH (in arbitrary units) of the 661.6 keV 
$^{137}$Cs peak acquired with the SensL SSPM + CsI(Tl) scintillation detector 
as a function of the temperature when the contribution of CsI(Tl) has been 
subtracted out and the change in the amplifier gain between the 3 temperature 
ranges has been compensated for. A normalization has also been performed so 
that the PH at $25\,^{\circ}\mathrm{C}$ is 1.}
\label{sensLSPMonly}
\end{figure}

The MPPC and its load resistor were both set on a breadboard, on which the 
4 signals from the MPPC dies were summed into one output which was
pre-amplified using a Hamamatsu H4083 photodiode charge-sensitive 
pre-amplifier board, also installed on the breadboard. This arrangement 
is shown on the top-right and bottom-right photos of Figure \ref{setup}.

The MPPC is blue-enhanced with a PDE which peaks at about 420 nm. We 
therefore chose to couple it to an encapsulated cylindrical 
(6 mm diameter x 6 mm long) 
Saint-Gobain Crystals' BrilLanCe 380 (LaBr$_{3}$:Ce; abreviated B380 on the 
Figures) scintillator using silicone grease. 
The PDE of the MPPC at the BrilLanCe 380 wavelength of maximum emission of 380 nm, when
operated at the recommended 70 V at $25\,^{\circ}\mathrm{C}$, is
about 23\% according to Hamamatsu. The hygroscopic crystal was hermetically 
sealed in an aluminum enclosure with a 5 mm thick optical exit made of glass. 
We placed a radioactive source (e.g. $^{137}$Cs or $^{22}$Na) directly 
on top of the breadboard next to our scintillation detector. The entire setup 
on the breadboard was placed in an aluminum can to ensure isolation from 
external noise and light-tightness. All external voltages were supplied through 
this aluminum can: we used one DC power supply to provide +12 V, -12V, and 
ground connections to the pre-amplifier board and another DC power 
supply (Tennelec TC 954) to provide the 65 V bias to the MPPC. The aluminum 
can itself was placed in the temperature test chamber specified in the 
previous Subsection.

\begin{figure*}
\centerline{
\mbox{\includegraphics[width=0.497\textwidth]{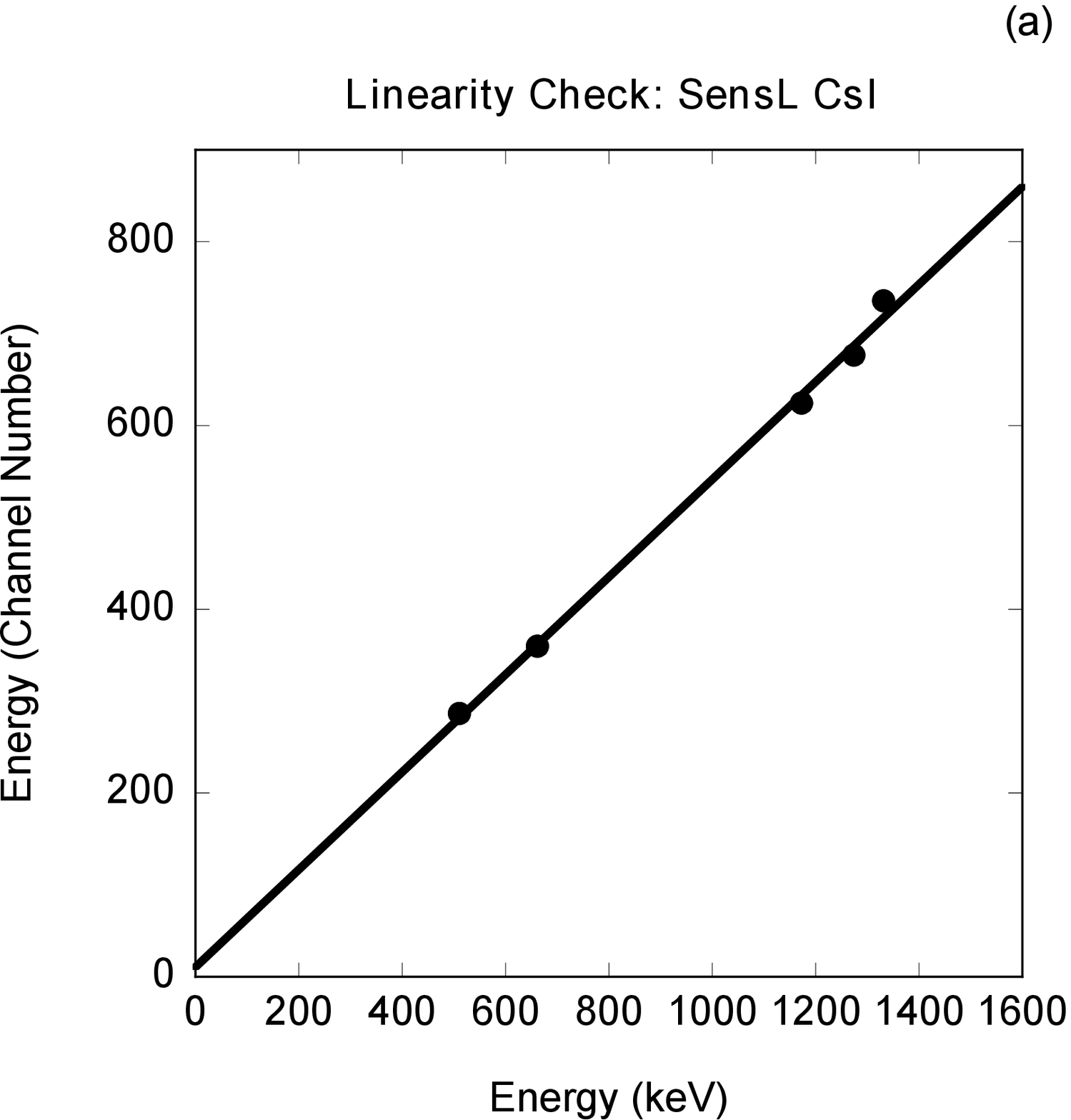}}
\mbox{\includegraphics[width=0.497\textwidth]{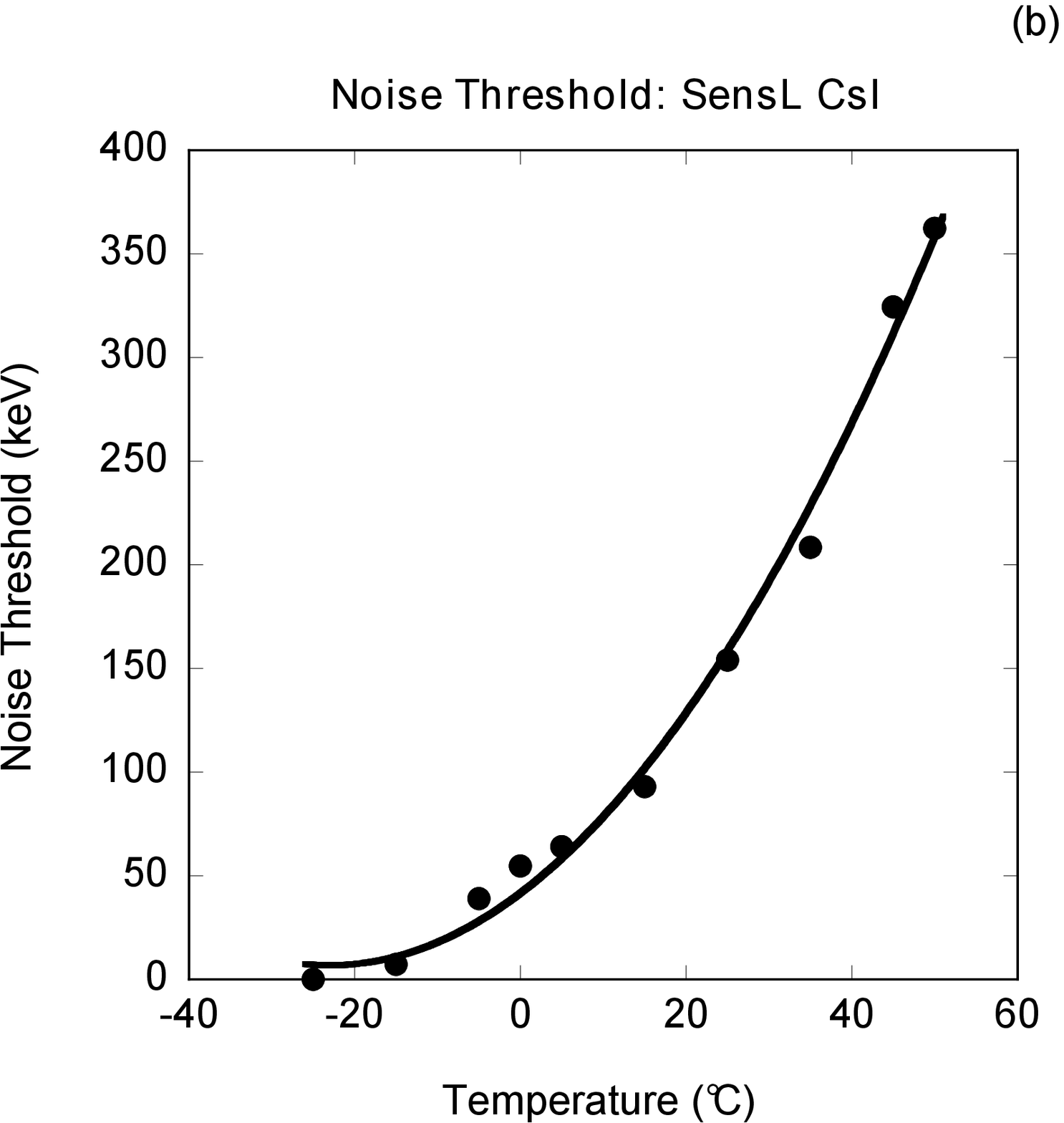}}
}

\caption{For the SensL SSPM + CsI(Tl) scintillation detector: (a) linearity 
check using lines in the $^{137}$ Cs, $^{22}$Na, and $^{60}$Co spectra,
the thick black line being a linear fit to the data points,  and 
(b) noise threshold versus temperature obtained using a $^{22}$Na source, 
where the thick black curve is the result of a quadratic fit to the data.}
\label{sensLlinearityNoise}
\end{figure*}

The pre-amplified signal of the summed MPPC output was further amplified by 
a Canberra 2020 spectroscopy amplifier (with a 250 ns shaping time, the 
smallest setting available on our amplifier module) and then digitized by 
a Canberra 8075 ADC module. The resulting output was analyzed and saved 
by an MCA.

\section{Results}

\subsection{Description of the measurements}

We studied the variations with temperature of the PH and PHR of the 661.6 keV 
peak in the $^{137}$Cs spectrum within the $-25\,^{\circ}\mathrm{C}$ - 
$+50\,^{\circ}\mathrm{C}$ range. Since the variations in PH were quite 
significant within this range, we broke it up into three sub-ranges:
$-25\,^{\circ}\mathrm{C}$ - $0\,^{\circ}\mathrm{C}$,
$0\,^{\circ}\mathrm{C}$ - $+25\,^{\circ}\mathrm{C}$, and
$+25\,^{\circ}\mathrm{C}$ - $+50\,^{\circ}\mathrm{C}$.
The amplifier gain setting was different for each of those 3 ranges. 

\begin{figure*}
\centering
\centerline{
\mbox{\includegraphics[width=0.497\textwidth]{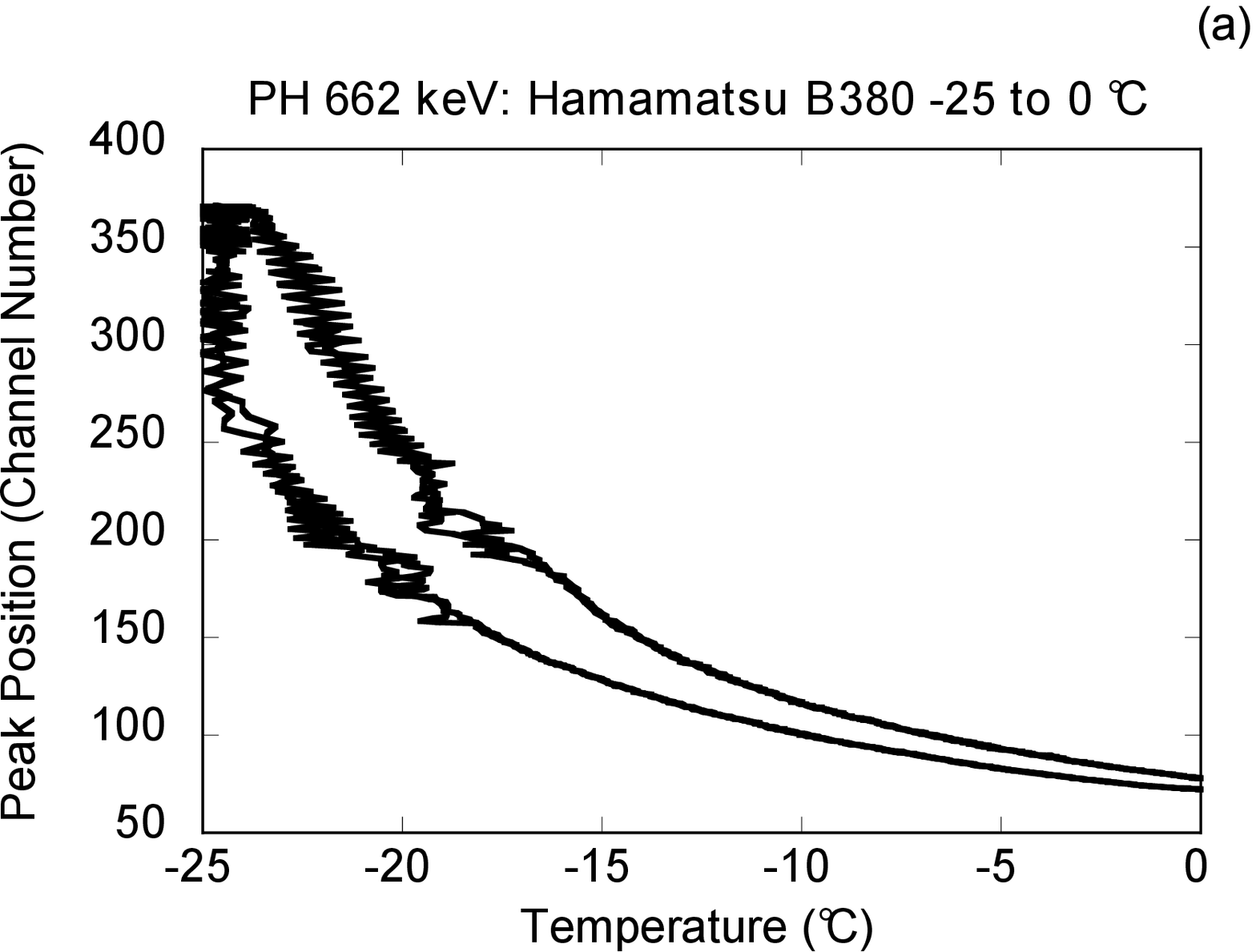}}
\mbox{\includegraphics[width=0.497\textwidth]{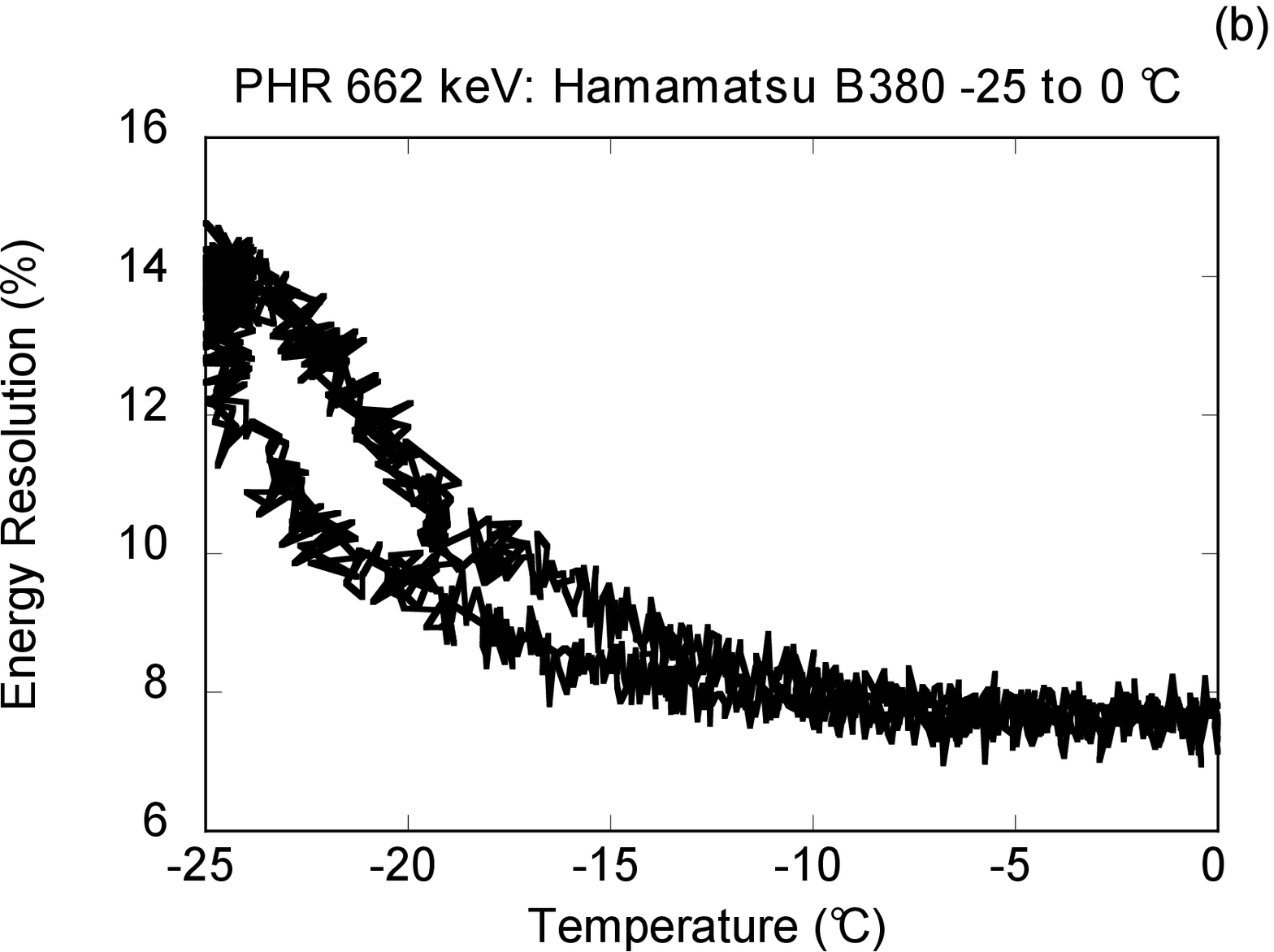}}
}
\centerline{
\mbox{\includegraphics[width=0.497\textwidth]{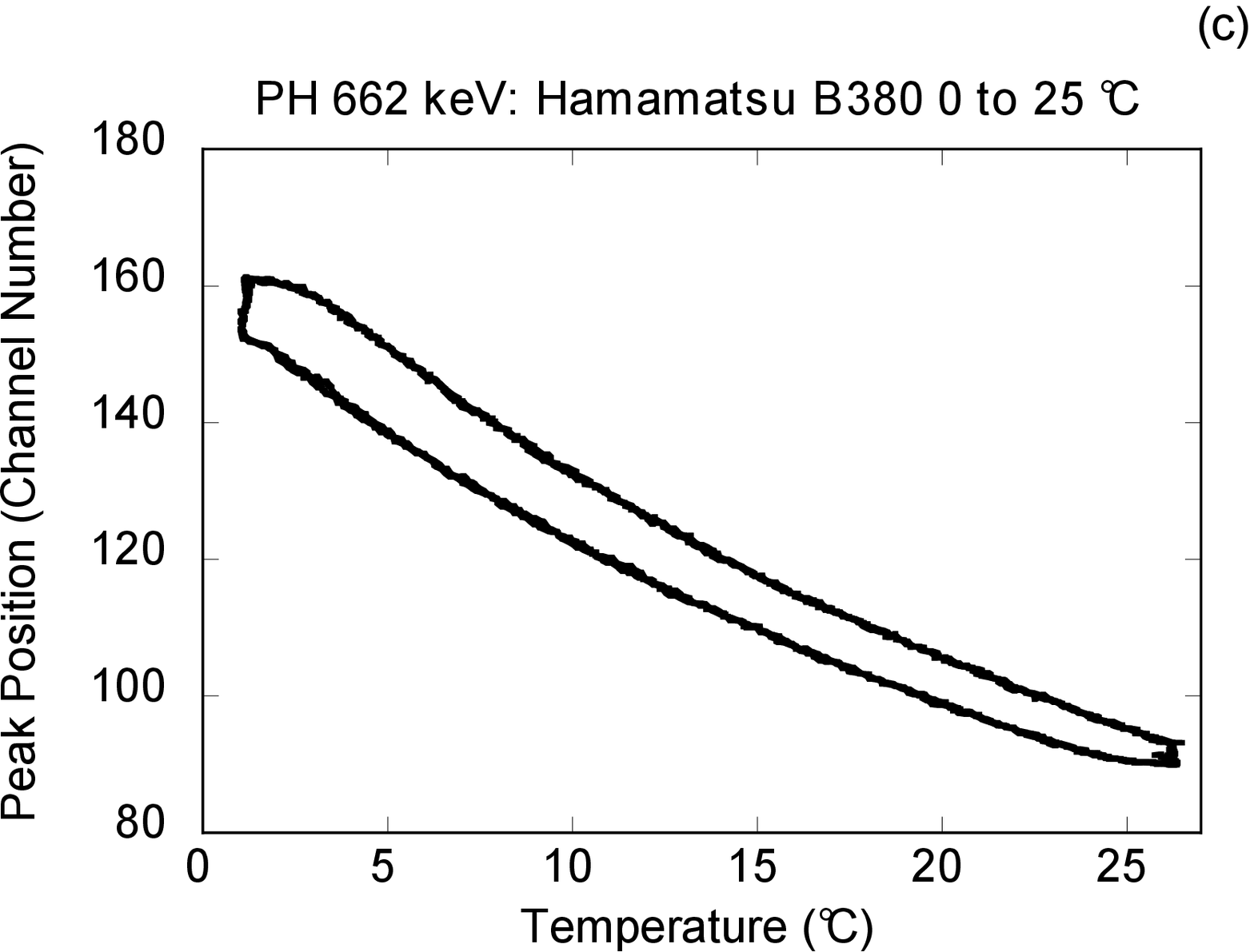}}
\mbox{\includegraphics[width=0.497\textwidth]{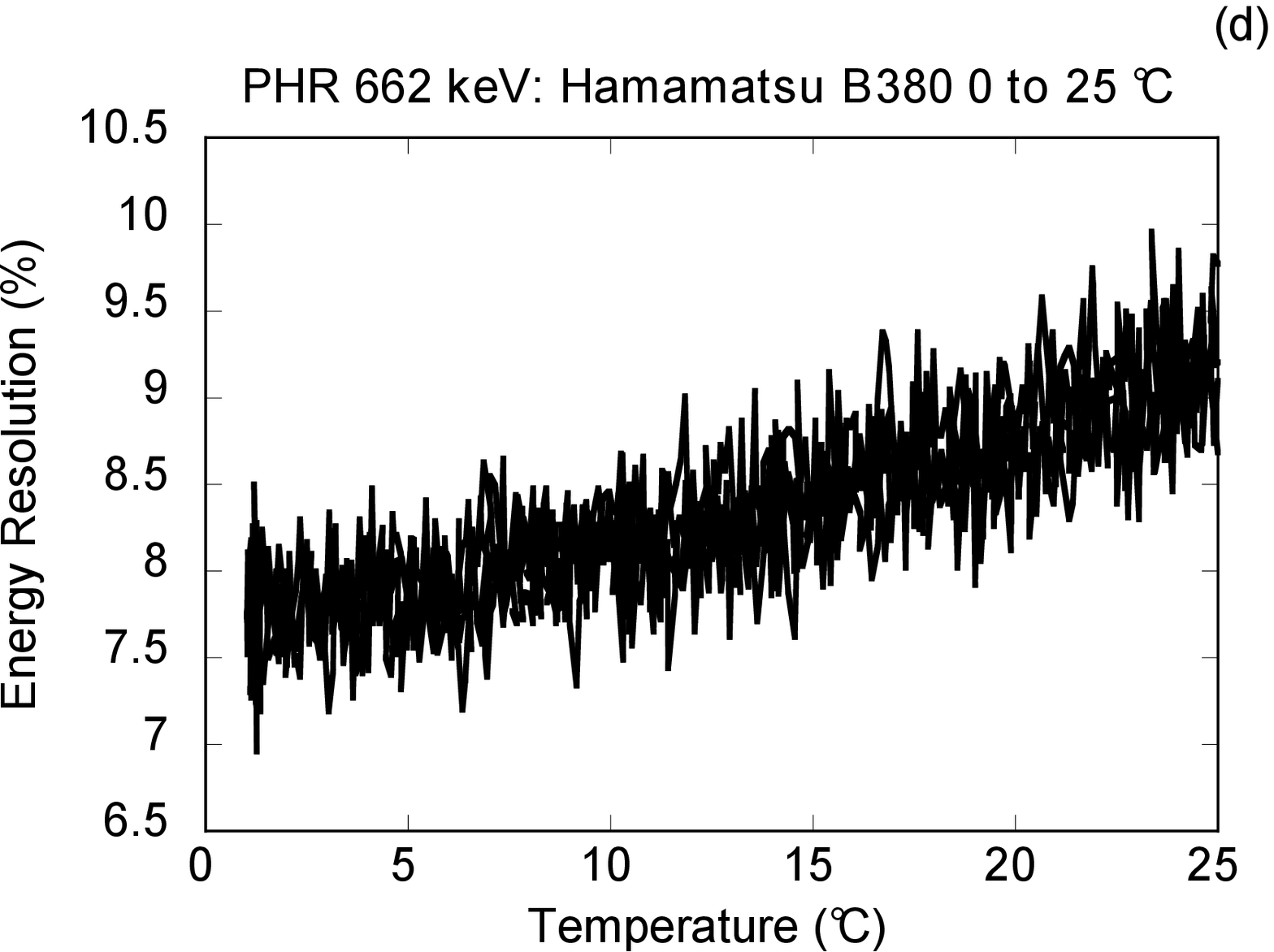}}
}
\centerline{
\mbox{\includegraphics[width=0.497\textwidth]{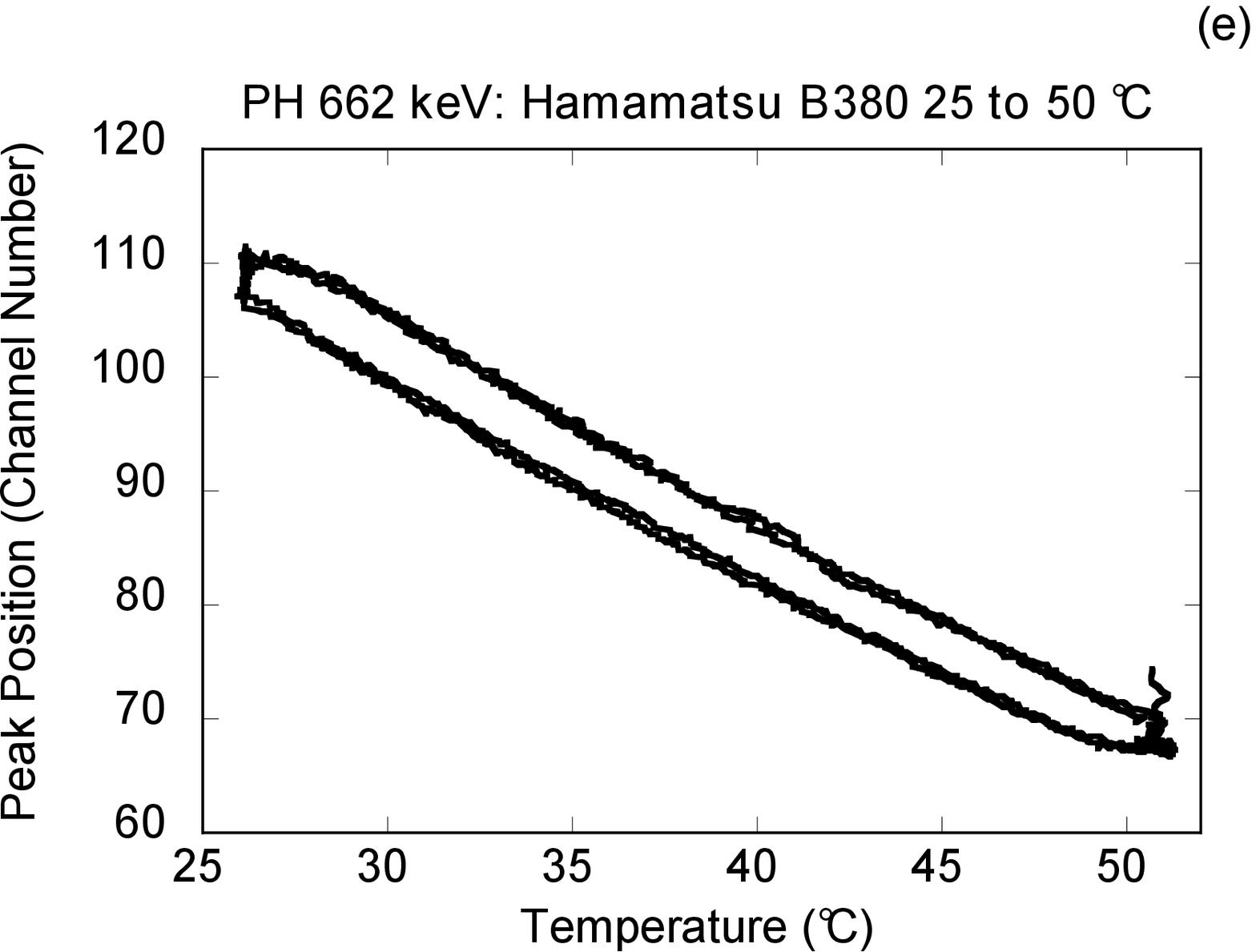}}
\mbox{\includegraphics[width=0.497\textwidth]{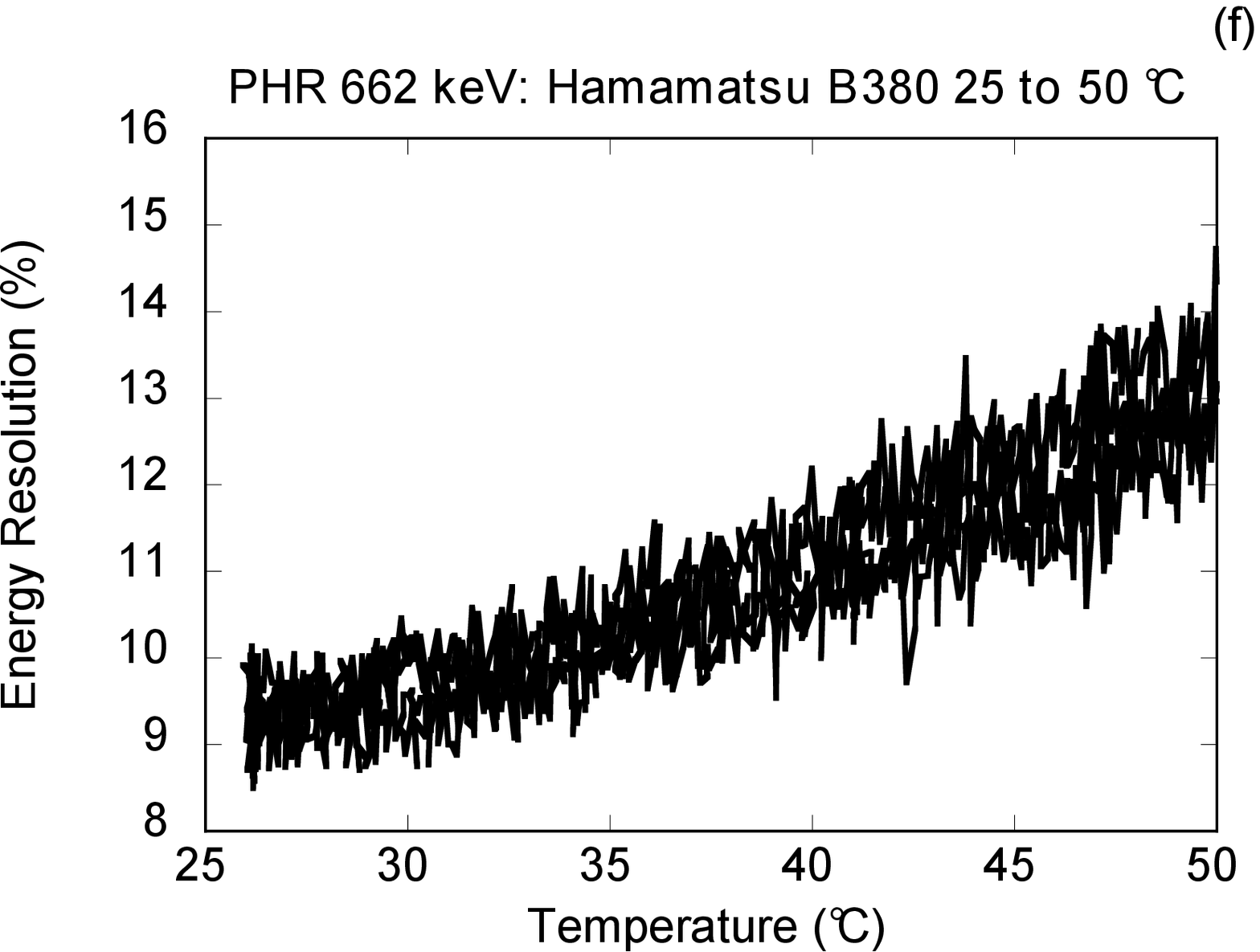}}
}
\caption{Temperature dependence, for the Hamamatsu MPPC + BrilLanCe 380 scintillation 
detector, of (a) the PH from -25 to 
$0\,^{\circ}\mathrm{C}$, (b) the PHR from -25 to $0\,^{\circ}\mathrm{C}$,
(c) the PH from 0 to $25\,^{\circ}\mathrm{C}$, (d) the PHR from
0 to $25\,^{\circ}\mathrm{C}$, (e) the PH from 25 to 
$50\,^{\circ}\mathrm{C}$, and (f) the PHR from 25 to 
$50\,^{\circ}\mathrm{C}$ of the 661.6 keV peak in the $^{137}$Cs spectrum.
The amplifier gain settings are different for each of the 3 temperature 
ranges.}
\label{hamamatsuTempDep}
\end{figure*}

\begin{figure}
\centering
\includegraphics[width=0.8\textwidth]{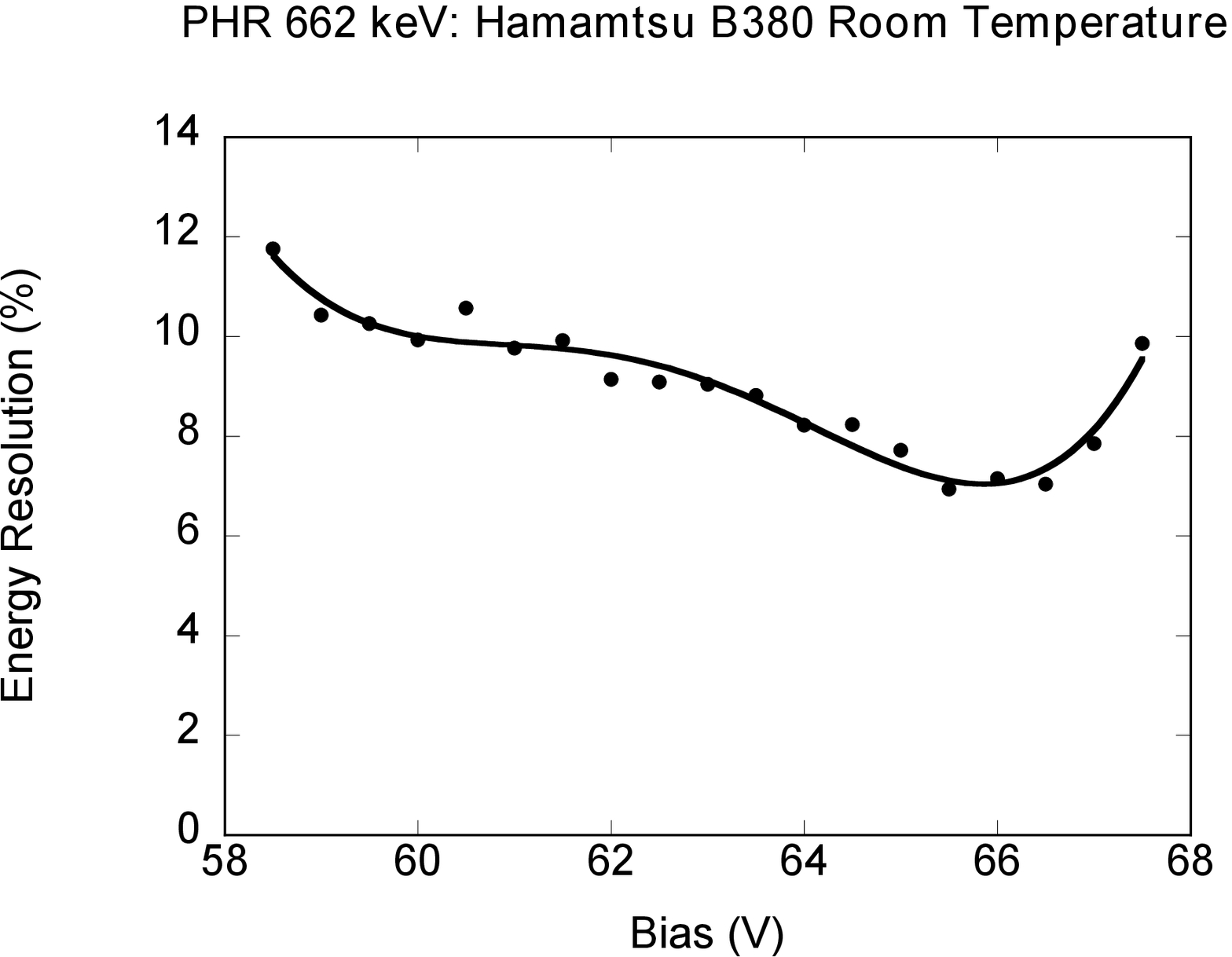}
\caption{PHR, for the Hamamatsu MPPC + BrilLanCe 380 scintillation detector, of the 
661.6 keV peak in the $^{137}$Cs spectrum as a function of bias at room 
temperature. The thick black curve is the result of a 
4$^{th}$-order polynomial fit to the data points}.

\label{hamBias}
\end{figure}

For each range, we began by soaking the detector at the highest temperature in 
that range for about 1 h and then adjusting the amplifier gain so that the 
661.6 keV peak was at the lowest end of the MCA display, since the PH is
expected to go up as the temperature is decreased. We then programmed the 
Thermotron temperature test chamber for a 28 h duration cycle: the
cycle begins with a 2 h  soak at the highest temperature in the range; then
the temperature is decreased by $25\,^{\circ}\mathrm{C}$ in 5 h (at the 
rate of $5\,^{\circ}\mathrm{C}$/h); this is followed by another 2 h soak 
at the newly reached temperature, the lowest in the cycle; then the 
temperature is ramped up again by $25\,^{\circ}\mathrm{C}$ in 5 h, at which
point the highest temperature in the range is reached again. This 14 h 
cycle is repeated one more time to complete the 28 h cycle for that 
particular range.  

\begin{table*}
\centering
\caption{Standard deviation (denoted by $\sigma$) and mean (denoted by $\mu$)
of the measured temperature during soaks 
at constant temperature for 4 temperature settings and the same statistics 
for the 661.6 keV PHR as measured by the Hamamatsu MPPC + BrilLanCe 380 scintillation
detector at those 4 temperatures. The statistics on the temperature are 
in $^{\circ}$C and those on the PHR in \%.}
\label{hamStats}
\begin{tabular*}{0.95\textwidth}{ccccc}
\hline 
\hline 
Temperature & $\mu$(Temperature) & $\sigma$(Temperature) 
& $\mu$(PHR) & $\sigma$(PHR)  \\
\hline 
-25  &  -24.42  & 0.46  & 13.64  & 0.68 \\
0    &    1.13  & 0.25  & 7.73  & 0.25 \\
25   &   26.14  & 0.24  & 9.34  & 0.31 \\
50   &   50.75  & 0.28  & 13.18  & 0.56 \\
\hline
\end{tabular*}
\end{table*}

\begin{figure}
\centering
\includegraphics[width=0.8\textwidth]{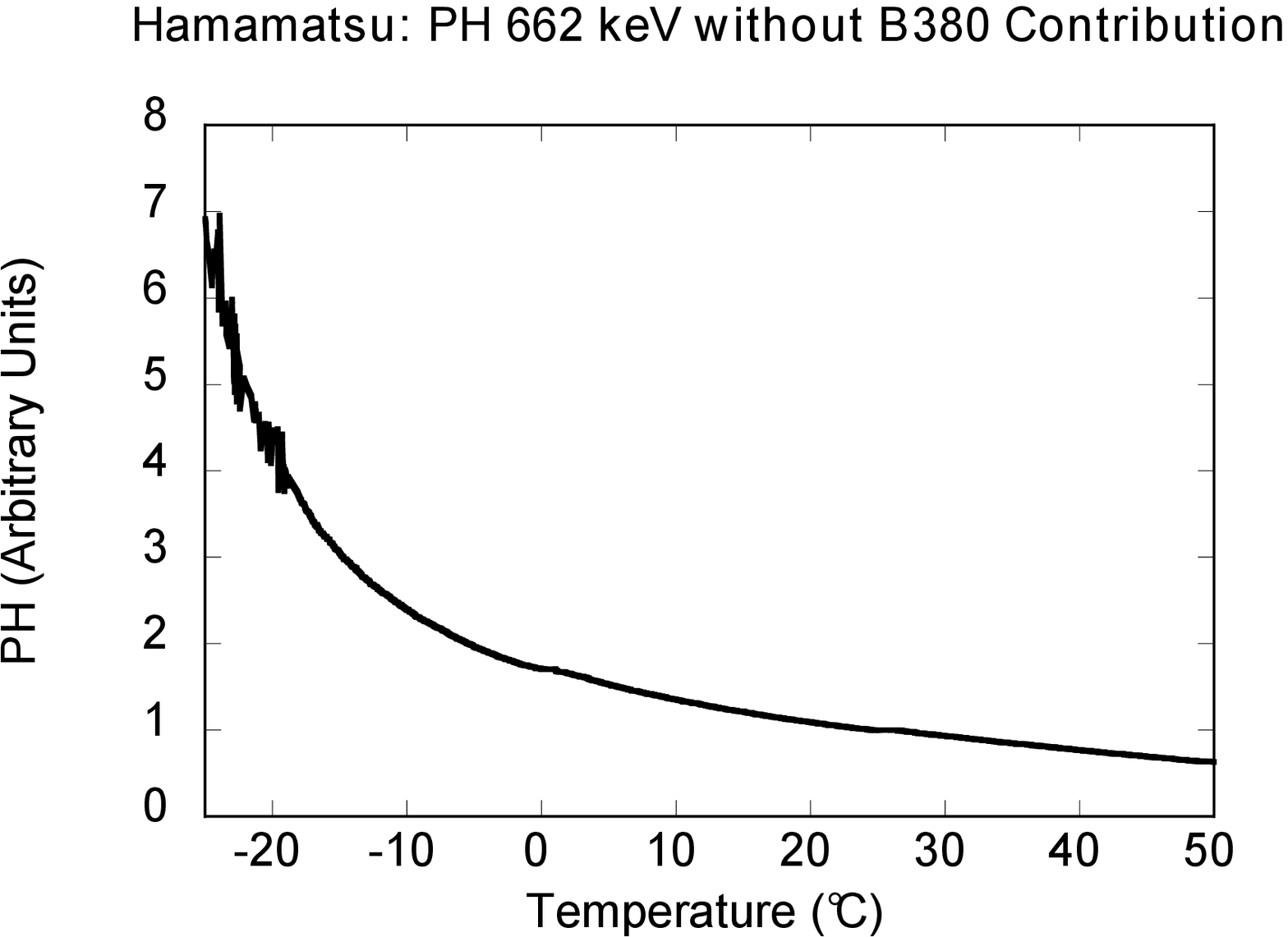}
\caption{PH (in arbitrary units) of the 661.6 keV 
$^{137}$Cs peak acquired with the Hamamatsu MPPC + BrilLanCe 380 scintillation detector 
as a function of the temperature when the contribution of the BrilLanCe 380 
crystal has been subtracted out and the change in the amplifier gain 
between the 3 temperature ranges has been compensated for. A normalization 
has also been performed so that the PH at $25\,^{\circ}\mathrm{C}$ is 1.}
\label{hamSPMonly}
\end{figure}

As the Thermotron chamber was cycling, data for a  $^{137}$ Cs spectrum was 
automatically accumulated for 60 s by the MCA and saved. As soon as one 
spectrum was saved, the next was started, with virtually no time gap between 
them. The temperature change within the 60 s acquisition time is negligibly 
small: about $0.08\,^{\circ}\mathrm{C}$. A record of the temperature was also 
made at the start of each acquisition period. More than 1,600 $^{137}$ Cs 
spectra were collected per temperature range and detector and then 
analyzed offline for PH and PHR.

\begin{figure*}
\centerline{
\mbox{\includegraphics[width=0.497\textwidth]{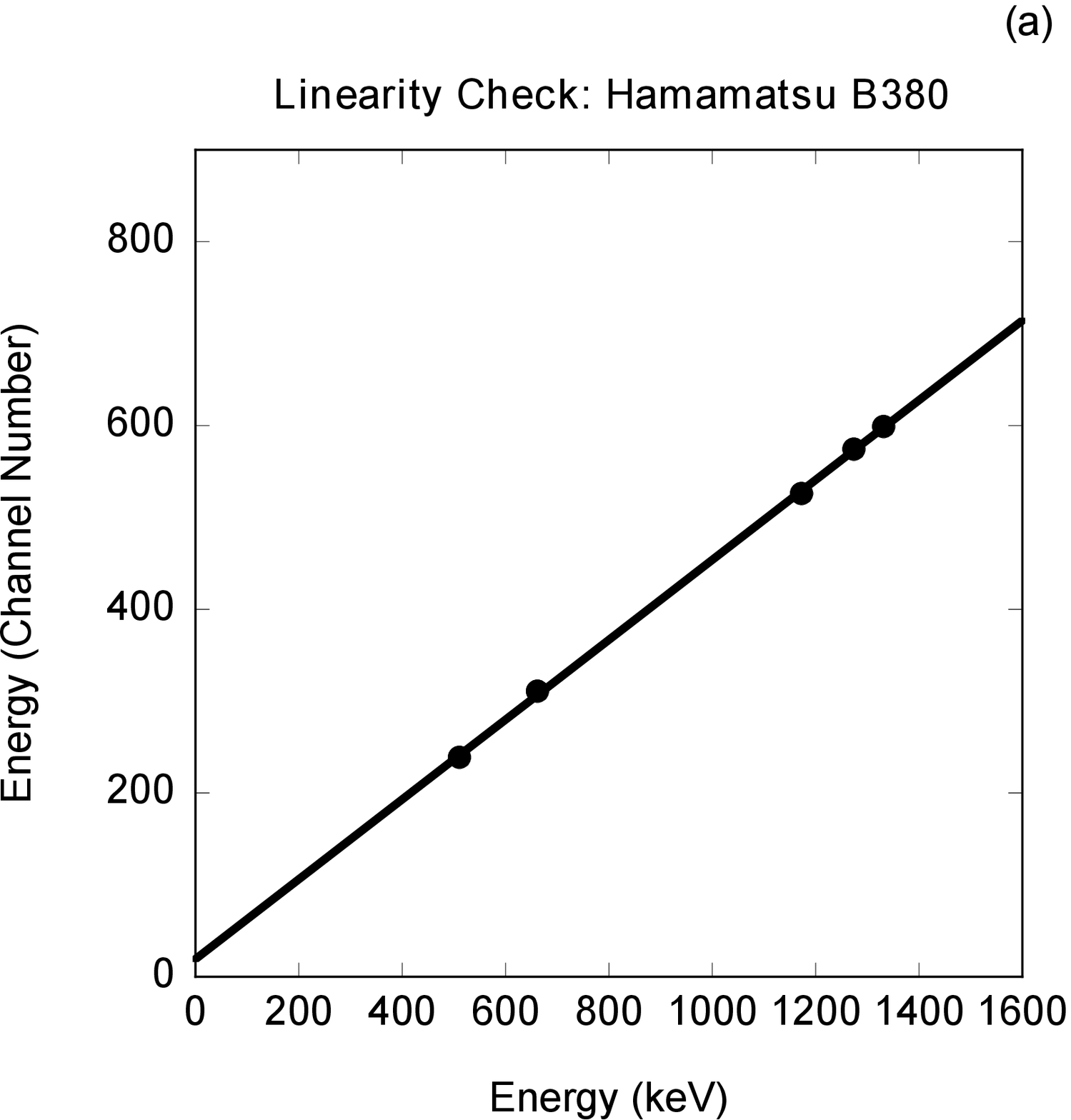}}
\mbox{\includegraphics[width=0.497\textwidth]{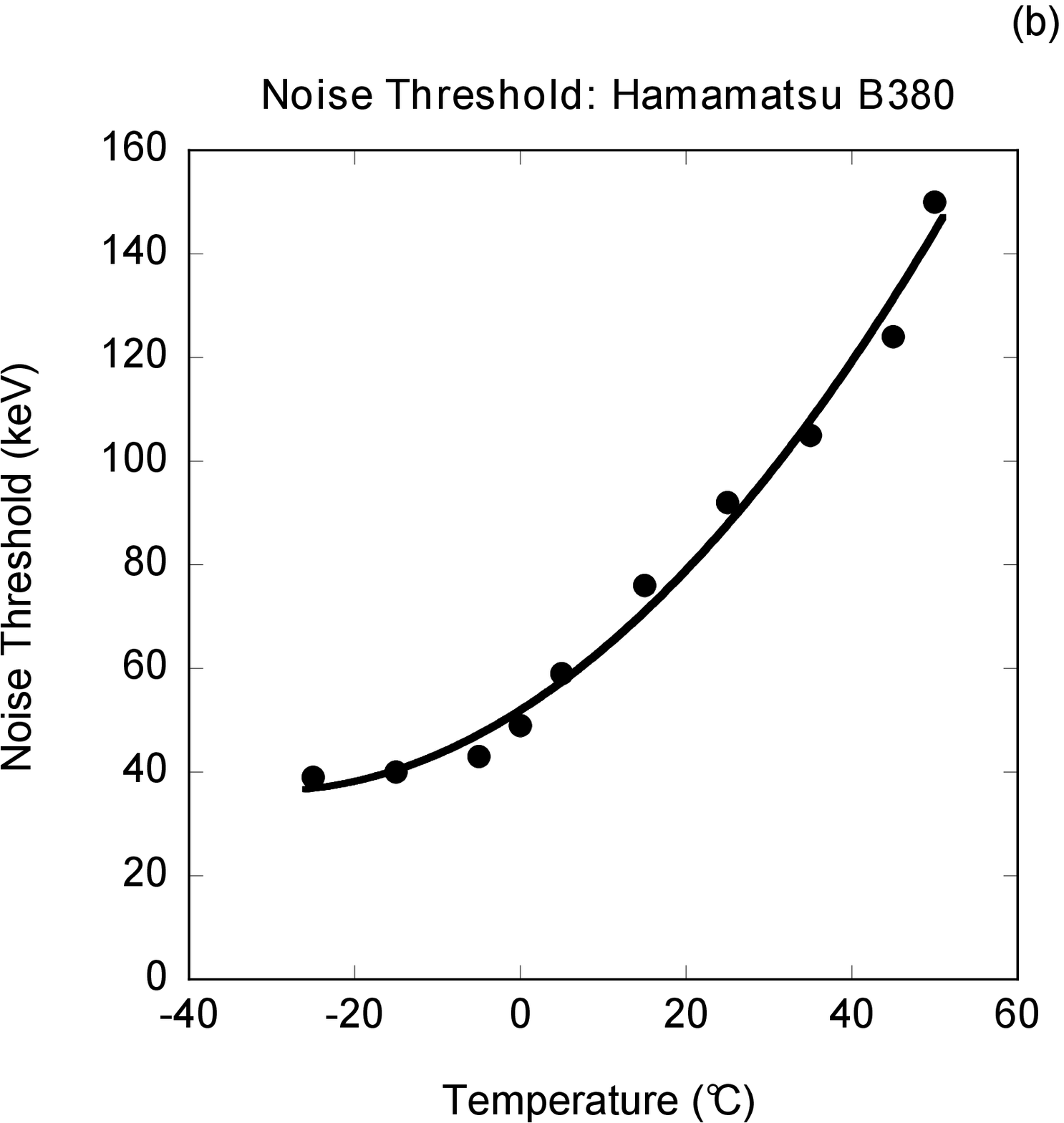}}
}

\caption{For the Hamamatsu MPPC + BrilLanCe 380 scintillation detector: (a) linearity 
check using lines in the $^{137}$ Cs, $^{22}$Na, and $^{60}$Co spectra, the
thick black line being a linear fit to the data points,  and 
(b) noise threshold versus temperature obtained using a $^{22}$Na source, where
the thick black curve is the result of a quadratic fit to the data.}
\label{hamLinearityNoise}
\end{figure*}

An energy linearity study was then made for each scintillation detector using a 
$^{137}$ Cs, a $^{22}$Na, and a $^{60}$Co source. Once the linearity was
confirmed up to 1332 keV, a noise-threshold study was performed
using the $^{22}$Na source, utilizing its 511 keV and 1274.5 keV lines 
for calibration. For these measurements, the Thermotron chamber was set 
to soak at the highest temperature ($50\,^{\circ}\mathrm{C}$) for 1 h; 50 min 
into the soak, a spectrum was acquired for typically 300 s which clearly 
showed the noise threshold and the two $^{22}$Na lines. Then the temperature 
was decreased (by typically $10\,^{\circ}\mathrm{C}$) in 15 min. The same 
procedure was repeated for a spectrum at this new temperature. 

\subsection{Results for the CsI(Tl)/SensL SSPM scintillation detector}

\subsubsection{Temperature dependence of the pulse height and pulse height resolution of the 661.6 keV peak in $^{137}$Cs}

Figure \ref{sensLTempDep} shows the PH and PHR of the 661.6 keV peak in the 
$^{137}$Cs spectrum as a function of temperature for the three temperature 
ranges considered. It is important to note that the amplifier gain settings 
were different for each range. Overall the PH decreases as the 
temperature is increased, while the PHR gradually worsens. This
trend is seen to become stronger as one goes from the 
$-25\,^{\circ}\mathrm{C}$ - $0\,^{\circ}\mathrm{C}$ range to the
$0\,^{\circ}\mathrm{C}$ - $+25\,^{\circ}\mathrm{C}$, and then on to the 
$+25\,^{\circ}\mathrm{C}$ - $+50\,^{\circ}\mathrm{C}$ range. The greater
fluctuations in PHR at $+50\,^{\circ}\mathrm{C}$ (as seen on 
Figure \ref{sensLTempDep}(f)) are due to both lower statistics in the 661.6 keV 
peak and to increased fluctuations in temperature as shown in 
Table \ref{sensLstats}, which is due to a marginal insulation 
system where the cables exit the thermal chamber. The seemingly erratic 
jumps in both the PH and PHR for 
small changes in temperature are presumably due to the fact that the 
detector never quite gets a chance to stabilize at any given temperature 
even at the relatively slow temperature change rate of $5\,^{\circ}\mathrm{C}$
per hour.Figure \ref{fig2} also shows that the SSPM exhibits a slight
hysteresis: the PH evolves on 2 slightly different curves when one decreases 
the temperature from the highest temperature in the range to the lowest versus 
when one increases it from the lowest temperature in the range back to 
the highest.

In Figure \ref{sensLSPMonly}, the PH of the 661.6 keV peak
has been plotted against the temperature in arbitrary units in the 
entire $-25\,^{\circ}\mathrm{C}$ - $+50\,^{\circ}\mathrm{C}$ range, when the 
contribution of CsI(Tl) has been subtracted out and the change in the 
amplifier gain between the three temperature ranges has been compensated for. 
A normalization has also been performed so 
that the PH at $25\,^{\circ}\mathrm{C}$ is 1. The PH attributable to 
the SSPM is seen to vary by a factor of 16.1 between $-25\,^{\circ}\mathrm{C}$ 
and $+50\,^{\circ}\mathrm{C}$.

An explanation about how we subtracted out the crystal contribution is in 
order here: the CsI(Tl) contribution is a number between 0 and 1 and 
was obtained from data available in the Saint-Gobain Crystals' CsI(Tl) 
data sheet (see \cite{sgc09}). The data consists of a curve showing the 
crystal's light output for 661.6 keV gamma rays as a function of temperature, 
normalized to the crystal's light output at $25\,^{\circ}\mathrm{C}$. We 
derived a functional form for this curve by a least squares fit. We used
this functional form to calculate the crystal contribution for each temperature.
We then simply divided the full detector response at a given temperature 
by the crystal contribution at that temperature.

The best PHR at 661.6 keV which we observed at room temperature was about 
12\% FWHM. When we coupled the same CsI(Tl) crystal to a Hamamatsu R1306-01 PMT
(using silicone grease), the PHR at 661.6 keV was 7.9\%.

\subsubsection{Linearity and temperature dependence of the noise threshold}

We checked the linearity of the detector using the 661.6 keV line in 
$^{137}$Cs, the 511 keV and 1274.5 keV lines in $^{22}$Na, and the 
1173 keV and 1332 keV lines in $^{60}$Co. The results are illustrated in Figure 
\ref{sensLlinearityNoise}(a) and show that the detector exhibits good 
linearity up to 1332 keV.

Once the linearity was checked, we performed a study of the 
noise threshold in the
$^{22}$Na spectrum as a function of temperature, using its 511 keV and 1274.5
keV lines for calibration. The threshold is determined by the intersection of
a linear fit to the steep fall-off of the noise edge and a linear fit to the 
relatively flat Compton continuum in the low-energy region. 
Figure \ref{sensLlinearityNoise}(b) shows that 
this threshold quadratically increases from essentially 0 keV at
$-25\,^{\circ}\mathrm{C}$ to about 362 keV at $+50\,^{\circ}\mathrm{C}$. It is 
worth noting the significant deterioration in performance in terms of 
noise levels near $+50\,^{\circ}\mathrm{C}$. 

\subsection{Results for the BrilLanCe 380/Hamamatsu MPPC scintillation detector}

\subsubsection{Temperature dependence of the pulse height and pulse height resolution of the 661.6 keV peak in $^{137}$Cs}

Figure \ref{hamamatsuTempDep} shows the variations in the PH and PHR of the 
661.6 keV peak in the $^{137}$Cs spectrum as a function of temperature for 
the three temperature ranges studied. The PH decreases as the temperature is 
increased, as expected. The PHR, however, has the expected 
behavior only in the $0\,^{\circ}\mathrm{C}$ - $+50\,^{\circ}\mathrm{C}$ range,
as it worsens when the temperature is increased, while in the 
$-25\,^{\circ}\mathrm{C}$ - $0\,^{\circ}\mathrm{C}$ range, the PHR improves
as the temperature is increased, the opposite of what is expected and 
observed for the SensL SSPM. It is useful to note here a difference in the 
room temperature behavior of the SensL and Hamamatsu sensors which we observed:
while the PHR of 661.6 keV peak as a function of bias voltage at room 
temperature as measured by the SensL SSPM-based scintillation detector
is mostly constant over much of the recommended bias range (i.e. 
within 29 V and 32 V), the behavior of Hamamatsu MPPC-based detector 
appears to be more complicated as illustrated in Figure \ref{hamBias}. 
As the temperature is changed, V$_{br}$ changes. Since V$_{op}$ is held 
constant, this results in changes in the excess voltage V$_{ex}$, which 
explains much of the changes in both the PH and PHR observed. This 
situation is similar to one where V$_{br}$ is held constant while V$_{op}$
 is changed in the opposite direction. Otherwise stated, changing the 
temperature is somewhat equivalent to going up or down the curve in 
Figure \ref{hamBias}. Presumably, the more complicated shape of this 
curve is responsible for the difference in the PHR behavior observed 
for the Hamamatsu MPPC-based detector compared the one using the SensL SSPM.
The hysteresis, which we observed for the SensL SSPM, is manifest even more 
clearly in the PH and PHR results for the Hamamatsu SSPM as seen on 
Figure \ref{fig5}. This phenomenon seems to be an intrinsic feature of 
these sensors.

We also chose to operate the MPPC at a bias of 65 V based on 
Figure \ref{hamBias}. A bias of about 66.5 V gives the best PHR at 
room temperature. However, since moving the temperature is equivalent to 
moving along the PHR versus bias curve, we chose to avoid getting too 
close to any of the extremities of this curve as we change the temperature, 
at the expense of a somewhat worse resolution at room temperature.

The large fluctuations in temperature at both $-25\,^{\circ}\mathrm{C}$ and 
$+50\,^{\circ}\mathrm{C}$ are clearly visible on Figure \ref{hamamatsuTempDep}
and Table \ref{hamStats}. They are bigger in magnitude than for the SensL 
setup since we used one particularly thick cable for the Hamamatsu setup
(the one used to supply +12 V, -12V, and the ground connections to the 
pre-amplifier board through the Aluminum can). The use of this thick cable 
meant a larger hole through the foam-stuffed exit window of the thermal chamber
to reach the voltage supply located outside of the chamber. Hence it was more 
difficult to achieve good thermal insulation, especially at the extreme 
temperatures of our range. 

Figure \ref{hamSPMonly} shows the PH of the 661.6 keV peak versus 
the temperature in arbitrary units in the 
entire $-25\,^{\circ}\mathrm{C}$ - $+50\,^{\circ}\mathrm{C}$ range, when the 
contribution of BrilLanCe 380 has been subtracted out and the change in the 
amplifier gain between the three temperature ranges has been compensated for. 
A normalization has also been performed so 
that the PH at $25\,^{\circ}\mathrm{C}$ is 1. The BrilLanCe 380 contribution was 
obtained from data available in the Saint-Gobain Crystals' BrilLanCe 380 data 
sheet (see \cite{sgc09}). The PH varies by a factor of 11.2 between 
$-25\,^{\circ}\mathrm{C}$ and $+50\,^{\circ}\mathrm{C}$.

The best PHR at 661.6 keV which we observed at room temperature was 6.9\%
(at a bias setting of 66.5 V). When we coupled the same BrilLanCe 380 crystal to a 
Hamamatsu R1306-01 PMT (with silicone grease), however, the PHR at 661.6 
keV was 3.5\%. 

\subsubsection{Linearity and temperature dependence of the noise threshold}

 We used the 661.6 keV line in $^{137}$Cs, the 511 keV and 1274.5 keV 
lines in $^{22}$Na, and the 1173 keV and 1332 keV lines in $^{60}$Co to check 
the linearity of the scintillation detector. The results are illustrated in 
Figure \ref{hamLinearityNoise}(a) and show that the detector exhibits good 
linearity up to 1332 keV.

Once the linearity checked, we performed a study of the noise threshold in the
$^{22}$Na spectrum as a function of temperature, using its 511 keV and 1274.5
keV lines for calibration. Figure \ref{hamLinearityNoise}(b) shows that 
this threshold quadratically increases from 39 keV at
$-25\,^{\circ}\mathrm{C}$ to about 150 keV at $+50\,^{\circ}\mathrm{C}$. The 
noise level is worse than for SensL at $-25\,^{\circ}\mathrm{C}$, but much
better at $+50\,^{\circ}\mathrm{C}$. The fact that the noise threshold at 
$-25\,^{\circ}\mathrm{C}$ is not zero as it was for the SensL SSPM may be due 
to the fact that our alumininum can provides worse isolation from external 
noise than the SensL pre-amplifier/power board.

\section{Discussion}

Our measurements show a very strong dependence of the PH of 661.6 keV gamma
rays on the temperature due to drifts in V$_{br}$, and hence in the SSPM gain
via V$_{ex}$. When the contribution of the crystal is subtracted out, the 
PH varies by a factor of 16.1 and 11.2 for the SensL and Hamamatsu SSPM-based
detectors, respectively, when the temperature is raised from 
$-25\,^{\circ}\mathrm{C}$ to $+50\,^{\circ}\mathrm{C}$. The slope, 
at $+25\,^{\circ}\mathrm{C}$, of the curves
showing the PH at 661.6 keV as a function of temperature is typically a 
few percent per degree Celcius, which is about the same as what has been 
reported 
for an avalanche photodiode (\cite{fla09}). PMTs also have a gain that
varies with temperature, but to a lesser degree for typical bi-alkali PMTs 
($< 0.1$\% per degree Celcius; see \cite{ham09}). 
As with PMT systems, but to a greater degree, a SSPM used in an application
where the temperature will vary, would require a circuit to monitor 
temperature or gain, and feedback to readjust V$_{op}$ to hold V$_{ex}$ and
the overall gain constant.

Gain stabilization will not affect noise edge: the noise threshold for 
the SensL SSPM-based detector is 
362 keV at $+50\,^{\circ}\mathrm{C}$, while for the Hamamatsu MPPC-based 
detector, it is 150 keV at the same temperature.  A
cooling system, for example based on a Peltier module, could be used to 
maintain temperature and performance roughly constant with some increase
in overall size and power consumption.

Finally, both SSPMs tested underperformed in terms of PHR at room temperature
compared to a regular PMT; the best PHR we observed for our CsI(Tl) crystal
using the SensL SSPM at room temperature was about 12\%, while we were able 
to achieve 7.9\% with a Hamamatsu R1306-01 PMT coupled to the same 
crystal using silicone grease. The best room temperature PHR which we observed
for the BrilLanCe 380 crystal using the Hamamatsu MPPC was 6.9\%, 
while the same crystal, when coupled with silicone grease to a Hamamatsu 
R1306-01 PMT, yielded 3.5\%.

\section{Conclusion}

We have studied the variations in the PH and PHR of the 661.6 keV gamma
rays in $^{137}$Cs as a function of temperature within the
$-25\,^{\circ}\mathrm{C}$ - $+50\,^{\circ}\mathrm{C}$ range using 2 
state-of-the-art SSPM-based scintillation detectors: one where a 
13 mm x 13 mm, 16-die SSPM by SensL was coupled to 
a 13 mm x 13 mm x 13 mm CsI(Tl) crystal by Saint-Gobain Crystals; one 
where a 6 mm x 6 mm, 4-die MPPC by Hamamatsu was coupled to a
6 mm (diameter) x 6 mm (length) BrilLanCe 380 (LaBr$_{3}$:Ce) crystal also by
Saint-Gobain Crystals. We also measured the noise threshold in the 
$^{22}$Na spectrum for both detectors  as a function of the temperature within
the same range and checked their linearity at room temperature. 

Both detectors exhibited good linearity up to 1332 keV. There is a 
strong dependence of the PH and PHR on the 
temperature: while the PH for 661.6 keV gamma rays varied by a factor of 16.1 
in the temperature range considered using the SensL SSPM-based detector, 
it varied by a factor of 11.2 for the Hamamatsu MPPC-based detector in the 
same range. PHR also increased with temperature for both devices, but a 
direct comparison is not possible because of the difference in their design 
as well as the fact they were tested with different crystals. At
$+50\,^{\circ}\mathrm{C}$, the SensL device with CsI(Tl) gave 32.1\% FWHM while 
the best room temperature value observed was about 12\%. At 
$+50\,^{\circ}\mathrm{C}$, the Hamamatsu sensor registered 13.2\% with 
LaBr${3}$:Ce and at best 6.9\% at room temperature. The same two crystals
measured with a Hamamatsu R1306-01 PMT showed 7.9\% (CsI(Tl))and 3.5\% 
(LaBr${3}$:Ce) at room temperature.  
Noise edges rise with temperature reaching 362 keV and 
150 keV at $+50\,^{\circ}\mathrm{C}$ for the SensL and Hamamatsu SSPM-based 
detectors, respectively.

\section{Acknowledgements}

We wish to acknowledge the skilled assistance of Renee Gaspar and Brian
Bacon in preparing the CsI(Tl) crystal. We also wish to tend our appreciation
to Michael R. Mayhugh and Peter Menge for valuable discussions and 
insight and to Hamamatsu and SensL for providing us with the SSPMs used in 
the experiments. Our measurements have been performed using the Saint-Gobain 
Crystals test facilities.

\end{document}